\documentclass{article}
\usepackage{graphicx}
\usepackage{cite}
\usepackage{epsfig}
\usepackage{amsfonts}
 \usepackage{amssymb}
\usepackage{amsmath}
\usepackage{slashed}

\usepackage{hyperref}
\usepackage{graphicx}       % Include figure files
\usepackage{dcolumn}        % Align table columns on decimal point
\usepackage{bm}            		 % bold math
\usepackage{amssymb}
\usepackage{amstext}
\usepackage{tensor}
\usepackage{color} 
\usepackage{transparent}

\date{\today}

\newcommand{\insertplot}[5]{\begin{figure}
 \hfill\hbox to 0.05in{\vbox to #5in{\vfill
 \inputplot{#1}{#4}{#5}}\hfill}
 \hfill\vspace{-.1in}
 \caption{#2}\label{#3}
 \end{figure}}
 \newcommand{\inputplot}[3]{
 \special{ps: plotfile #1}
}
\newcounter{fig}

\newcommand{\ee}{\end{equation}}
\newcommand{\eea}{\end{eqnarray}}
\newcommand{\bea}{\begin{eqnarray}}

\newcommand{\beq}{\begin{equation}}
\newcommand{\eeq}{\end{equation}}

\newcommand{\ze}{\kern 0.05em}

\begin{document}
\begin{center}

{\Large \bf Spontaneous scalarisation of charged black holes: \\ coupling dependence and dynamical features}
\vspace{0.8cm}
\\
{Pedro G. S. Fernandes$^{\dagger}$,  
Carlos A. R. Herdeiro$^{\dagger}$,  
Alexandre M. Pombo$^{\ddagger \Diamond}$, \\
Eugen Radu$^{\ddagger  \Diamond}$ and
Nicolas Sanchis-Gual$^{\dagger}$
\vspace{0.3cm}
\\
$^{\dagger}${\small Centro de Astrof\'\i sica e Gravita\c c\~ao - CENTRA,} \\ {\small Departamento de F\'\i sica,
Instituto Superior T\'ecnico - IST, Universidade de Lisboa - UL,} \\ {\small Avenida
Rovisco Pais 1, 1049-001 Lisboa, Portugal}
\vspace{0.3cm}
\\
$^{\ddagger }${\small  Center for Research and Development in Mathematics and Applications (CIDMA),
\\
Campus de Santiago, 3810-183 Aveiro, Portugal}
\vspace{0.3cm}
\\
$^{\Diamond}${\small Departamento de F\'\i sica da Universidade de Aveiro,
Campus de Santiago, 3810-183 Aveiro, Portugal}}
\vspace{0.3cm}
\end{center}

 \date{January 2019}

	\begin{abstract}   
    Spontaneous scalarisation of electrically charged, asymptotically flat Reissner-Nordstr\"om black holes (BHs) has been recently demonstrated to occur in Einstein-Maxwell-Scalar (EMS) models. This phenomenon is allowed by a non-minimal coupling between the scalar and the Maxwell fields, and does not require non-minimal couplings of the scalar field to curvature invariants. EMS BH scalarisation presents a technical simplification over the BH scalarisation that has been conjectured to occur in extended Scalar-Tensor Gauss-Bonnet (eSTGB) models. It is then natural to ask: 1) how universal are the conclusions extracted from the EMS model? And 2) how much do these conclusions depend on the choice of the non-minimal coupling function?  Here we address these questions by performing a  comparative analysis of several different forms for the coupling function including: exponential, hyperbolic, power-law and a rational function (fraction) couplings. In all of them we obtain and study the domain of existence of fundamental, spherically symmetric, scalarised BHs and compute, in particular, their entropy. The latter shows that  scalarised EMS BHs are always entropically preferred over the RN BHs with the same total charge to mass ratio $q$. This contrasts with the case of eSTGB, where for the same power-law coupling the spherical, fundamental scalarised BHs are not entropically preferred over the Schwarzschild solution. Also, while the scalarised solutions in the EMS model for the exponential, hyperbolic and power-law coupling are very similar, the rational function coupling leads to a transition in the domain of existence, by virtue of a pole in the coupling function, into a region of ``exotic" solutions  that violate the weak energy condition. Furthermore, fully non-linear dynamical evolutions of unstable RN BHs with different values of $q$ are presented. These show: 1) for sufficiently small $q$, scalarised solutions with (approximately) the same $q$ form dynamically; 2) for large $q$, spontaneous scalarisation visibly decreases $q$; thus evolutions are non-conservative; 3) despite the existence of non-spherical, static scalarised solutions, the evolution of unstable RN BHs under non-spherical perturbations leads to a spherical scalarised BH. 
	\end{abstract}

%
%%%%%%%%%%%%%%%%%%%%%%%%%%%%%%%%%%%%%%%%%%%%%%%%%%%%%%%%%%%%%%%%%%%%%%%%%%%%%%
%%%%%%%%%%%%%%%%%%%%%%%%%%%%%%%%%%%%%%%%%%%%%%%%%%%%%%%%%%%%%%%%%%%%%%%%%%%%%%
 \section{Introduction}\label{S1}
%%%%%%%%%%%%%%%%%%%%%%%%%%%%%%%%%%%%%%%%%%%%%%%%%%%%%%%%%%%%%%%%%%%%%%%%%%%%%%
%%%%%%%%%%%%%%%%%%%%%%%%%%%%%%%%%%%%%%%%%%%%%%%%%%%%%%%%%%%%%%%%%%%%%%%%%%%%%%
%
Never, in their one hundred years of history, there has been a more exciting time to study black holes (BHs). A diversity of observational data is delivering information with unprecedented accuracy on the strong gravity region around these objects - see $e.g.$ the reviews~\cite{Berti:2015itd,Barack:2018yly}. These data include, in particular, the gravitational waves events that have been observed as a result of BH binaries inspiral and merger,  initiated with the epoch-making detection of the first transient, \textsc{gw150914}~\cite{Abbott:2016blz}; the catalogue of gravitational wave events, as of February 2019, is given in~\cite{LIGOScientific:2018mvr}. Another exciting piece of  observational evidence comes from  the near future release of the first image of a BH shadow by the Event Horizon Telescope collaboration~\cite{EHT} - see $e.g.$~\cite{Cunha:2018acu,Psaltis:2018xkc} for recent reviews on BH shadows. 

BHs have a surprisingly small number of macroscopic degrees of freedom in General Relativity (GR) \textit{and} electro-vacuum, where a remarkable uniqueness holds - see $e.g.$~\cite{Chrusciel:2012jk} for a review. In this framework, the only physical BH solution (with a connected event horizon) is the Kerr-Newman BH~\cite{Newman:1965my}, and astrophysically, only the zero charge limit (the Kerr BH~\cite{Kerr:1963ud}) is likely to be relevant. The Kerr solution has only two macroscopic degrees of freedom, and BHs in GR (\textit{and} electro-vacuum) are thus colloquially described as having ``no-hair"~\cite{ruffini1971introducing}. Gravitational theories beyond GR or even GR with matter sources ($i.e.$ beyond electro-vacuum) allow a much richer landscape of BH solutions - see $e.g.$ the reviews~\cite{herdeiro2015asymptotically,Volkov:2016ehx} for different types of non-Kerr BHs. These are often called ``hairy" BHs since they have more macroscopic degrees of freedom. Then, the central question becomes if there are~\textit{dynamically} viable ``hairy" BHs that could represent alternatives to the Kerr BH paradigm.

A dynamical mechanism that could lead to the formation of BHs that differ from the standard GR electro-vacuum BHs is~\textit{spontaneous scalarisation}. This phenomenon was proposed in the context of neutron stars in scalar tensor models~\cite{Damour:1993hw} in the 1990s. In this context, the presence of non-conformally invariant matter (such as a neutron star) sources scalar field gradients due to the non-minimal coupling of the scalar field to the Ricci curvature. For a certain region within the domain of existence of (scalar-free) neutron stars, it becomes energetically favourable to scalarise. And in this domain, the tendency to scalarise can be seen from a perturbative instability of the scalar-free solutions against scalar perturbations. It turns out that BHs are immune to this tendency to scalarise because they are conformally invariant in scalar-tensor theories, as BH solutions in these theories, in general, coincide with the electro-vacuum solutions~\cite{Hawking:1972qk,Sotiriou:2011dz}. Thus, they do not source scalar field gradients and do not scalarise. But if the BHs would be surrounded by non-conformally invariant matter they should scalarise in a similar way, as suggested in~\cite{Cardoso:2013opa,Cardoso:2013fwa}. This sort of BH scalarisation was confirmed in a set of concrete field theory models in~\cite{Herdeiro:2019yjy}. A similar phenomenon of ``tensorization" (for neutron stars) was discussed in~\cite{Ramazanoglu:2017xbl}.

Instead of considering the traditional scalar-tensor models, the recent focus on BH spontaneous scalarisation --  triggered by the works~\cite{Doneva:2017bvd,Antoniou:2017acq,Silva:2017uqg} -- has been centred on \textit{extended Scalar-Tensor-Gauss-Bonnet (eSTGB) gravity}. Historically, gravitational models with Gauss-Bonnet (GB) curvature corrections have appeared in the context of Lovelock gravity~\cite{Lovelock:1971yv}, where the GB combination becomes dynamical in higher dimensions, or in the context of string theory, where the GB combination has been argued to arise naturally~\cite{Zwiebach:1985uq} and, if a dilatonic coupling is included, can become dynamical in four spacetime dimensions. BHs in the latter context have been first obtained in~\cite{Kanti:1995vq}. The Kerr family (including Schwarzschild) does not solve the corresponding equations of motion; new BH solutions appear which are perturbatively stable in some part of their domain of existence~\cite{Kanti:1997br}. 

 The class of models dubbed eSTGB gravity consist on allowing a more general coupling between the scalar field and the Gauss-Bonnet combination. If this coupling preserves a $\mathbb{Z}_2$ symmetry, then the model allows a scalar-free solution~\cite{Doneva:2017bvd,Silva:2017uqg}.  But the scalar-free solution seems to be generically unstable against scalar perturbations. Since, these models also allow the existence of scalarised BHs, the phenomenon of \textit{spontaneous scalarisation} has been conjectured to occur: for some range of mass (in terms of the GB coupling constant) a Schwarzschild BH becomes unstable and transfers some of its energy to a ``cloud" of scalar particles around it. In the case of the exponential-type coupling used in~\cite{Doneva:2017bvd} the scalarised BHs are entropically favoured and the fundamental branch of scalarised BHs contains perturbatively stable solutions against radial perturbations~\cite{Blazquez-Salcedo:2018jnn}. Then, the scalarised BHs could be the endpoints of the evolution of unstable Schwarzschild BHs. But in the case of the power-law coupling used in~\cite{Silva:2017uqg}, the scalarised solutions are not entropically favoured and the whole fundamental branch appears to be unstable against radial perturbations~\cite{Doneva:2017bvd}. In this case, therefore, it is unclear how the instability of the Schwarzschild solution terminates. See~\cite{Doneva:2018rou,Brihaye:2018bgc,Minamitsuji:2018xde,Silva:2018qhn,Bakopoulos:2018nui,Brihaye:2018grv,Cano:2019ore,Ramazanoglu:2019gbz} for additional recent work on eSTGB BH scalarisation.

In eSTGB gravity, spontaneous scalarisation is triggered by the strong spacetime curvature, which induces non-linear curvature terms in the evolution equations. These are computationally demanding and make dynamical studies challenging. As pointed out in \cite{Herdeiro:2018wub}, however, in what concerns the BH spontaneous scalarisation phenomenon, the eSTGB model belongs to a wider universality class that also contains the \textit{Einstein-Maxwell-Scalar} (EMS) models.  In these models, scalarisation occurs for electrically charged BHs and it is triggered by large enough charge to mass ratio, $q$. EMS theories have helped to gain a deeper insight into the  BH spontaneous scalarisation phenomena. This technically simpler model allowed an easier study of the domain of existence of solutions, in particular beyond the spherical sector, and it also allowed carrying out fully non-linear dynamical evolutions establishing that the instability of the scalar-free solution terminates in the scalarised BHs of the model~\cite{Herdeiro:2018wub}.  In this context,  the first examples of static, asymptotically flat, regular on and outside the event horizon BHs without spatial isometries have been constructed, but their dynamical role has been left unaddressed. Here we shall give evidence these solutions do not form dynamically, and are likely to be unstable. The fundamental, spherical, scalarised solutions, on the other hand, have been shown to be stable against generic perturbations (rather than only spherical)~\cite{Myung:2018jvi} - see also~\cite{Myung:2018vug,Boskovic:2018lkj} for additional work on related models. It is therefore relevant to ask how much the physics of the EMS and eSTGB models parallel each other, in what concerns the scalarisation phenomenon. Here we will point out that this parallelism depends on the choice of the coupling function. Moreover, we will also probe the dependence of the scalarised BHs of the EMS model on the choice of the coupling function that determines the non-minimal coupling between the scalar field and the Maxwell Lagrangian. 
Finally, several dynamical features, via fully non-linear numerical evolutions, will be pursued, in particular examining the constancy of the charge to mass ratio during the scalarisation process. We present evidence this is only approximately conserved for a sufficiently small value of $q$ of the initial, unstable RN BH.

This paper is organised as follows. We start, in Section~\ref{S2}, by presenting the basics of the EMS model and in particular specify possible coupling functions, that will be analysed in this work. These include: an exponential coupling, a hyperbolic ($\cosh$) coupling, a power-law coupling and a rational function (fractional) coupling. The numerical elliptic results are presented in Section~\ref{S3}, where the static solutions are obtained and the domain of existence discussed.  In particular, we show in more detail in Section~\ref{S4}, that for all examples of couplings considered the scalarised BHs are thermodynamically preferred over the electro-vacuum solutions -- the RN BHs with comparable global charges. In Section~\ref{S5} we address the time evolution problem and show that scalarised BHs do form dynamically, and compare the charge to mass ratio $q$ between the initial RN BH and the final scalarised BH. We also consider the evolution of unstable RN BHs under non-spherical perturbations to show that, in all cases, the end point is a spherically symmetric scalarised BH. Finally, in Section~\ref{S6} conclusions are presented.

%
%%%%%%%%%%%%%%%%%%%%%%%%%%%%%%%%%%%%%%%%%%%%%%%%%%%%%%%%%%%%%%%%%%%%%%%%%%%%%%
%%%%%%%%%%%%%%%%%%%%%%%%%%%%%%%%%%%%%%%%%%%%%%%%%%%%%%%%%%%%%%%%%%%%%%%%%%%%%%
 \section{The EMS models}\label{S2}
%%%%%%%%%%%%%%%%%%%%%%%%%%%%%%%%%%%%%%%%%%%%%%%%%%%%%%%%%%%%%%%%%%%%%%%%%%%%%%
%%%%%%%%%%%%%%%%%%%%%%%%%%%%%%%%%%%%%%%%%%%%%%%%%%%%%%%%%%%%%%%%%%%%%%%%%%%%%%
%
The EMS model describes a real scalar field $\phi $ minimally coupled to Einstein's gravity and non-minimally coupled to Maxwell's electromagnetism. The model is described by the action
	\begin{equation}\label{E21}
	 \mathcal{S}=\int d^4 x\ze \sqrt{-g}\ze\ze \left[ R - 
	2\ze \partial _\mu \phi\ze \partial ^{\ze \ze \mu} \phi - f_i \ze (\phi)\ze\ze \mathcal{I}\big( \psi,g\big) \right]\ ,
	\end{equation}
where  
$\mathcal{I}=F_{\mu \nu} \ze F^{\mu \nu}$ is the `source term' and $F_{\mu \nu}$ the usual Maxwell tensor.
 The coupling function $f_i\ze (\phi )$ couples the scalar field, non-minimally, to the Maxwell background; the subscript index $i$ will be used to label the various coupling choices, as specified below. 
The generic, spherically symmetric, line element which can be used to describe both a scalar-free and a scalarised BH solution is
	\begin{equation}
	\label{E22}
	 ds^2 = - N(r)\ze e^{-2\ze \delta (r)}dt^2+\frac{dr^2}{N(r)}+r^2 \big(d\theta ^2 +\sin ^2 \theta	\ze\ze	 d \varphi ^2\big)\ ,
	\end{equation}
	where $N(r)\equiv1-2\ze m(r) /r$, $m(r)$ is the Misner-Sharp mass function~\cite{Misner:1964je}. Spherical symmetry, in the absence of a magnetic charge, imposes an electrostatic $4-$vector potential, $A(r)=V(r)\ze dt$, and a scalar field solely radial dependent $\phi(r) $. 
	This allows us to define an effective Lagrangian from which the equations of motion can be derived as
	\begin{equation}\label{E23}
	 \mathcal{L}_{\rm eff} = 
	%\int d \Omega _2\ze\ze \mathcal{L} = 
	e^{-\delta} \ze m'-\frac{1}{2}\ze e^{-\delta} \ze r^2 N\ze \phi ^{'\ze 2}+\frac{1}{2} e^{\delta}\ze f_i (\phi)\ze r^2 V ^{'\ze 2}\ .
	\end{equation}
	Recall that the functions $m,\ \delta ,\ \phi ,\ V$ are all radially dependent only. 
	This dependence is from now on omitted for notation simplicity. The equations of motions are
	\begin{eqnarray}
	\label{E24}
	&&
	 m' = \frac{1}{2}\ze r^2 N \phi ^{'\ze 2}+\frac{1}{2}\ze e^{2\ze \delta}\ze f_i (\phi )\ze r^2 V ^{'\ze 2}\ ,
	%
	%\\ 
	\qquad \delta '= -r\ze \phi ^{'\ze 2}\ ,
	\label{E25}
	\\
	&&
	 \Big( e^{\delta}\ze f_i (\phi ) \ze r^2\ze V' \Big) ' = 0\ ,
	%\label{E26}
	%\\
	 \qquad \Big( e^{-\delta} r^2 N\ze \phi ' \Big) ' = -\frac{1}{2} 
	\dot f_i ^{\ze \ze }(\phi )\ze\ze e^{\delta }\ze\ze r^2\ze V ^{'\ze 2}\ ,
	\label{E27}
	\end{eqnarray}
where 
we denote $\dot f_i=df_i/d\phi$
(also 
$\ddot f_i=d^2f_i/d\phi^2$
),
while a prime denotes a derivative $w.r.t.$ the radial coordinate $r$.
	The equation  for the electric potential  yields the first integral 
	\begin{equation}\label{E28}
	 V' = -e^{-\delta }\frac{Q_e}{r^2 \ze f_i (\phi )},
	\end{equation}		
	the integration constant, $Q_e$, being the electric charge\footnote{
	After replacing the expression of the 1st integral \eqref{E28},
	the equations for mass functions and scalar field take the simpler form
		\begin{equation}
		\label{s1}
		m'=\frac{1}{2}\ze r^2 N \phi ^{'\ze 2}+\frac{Q_e^2}{2r^2 f_i(\phi)} \ , \qquad
		\phi''+\frac{1+N}{rN}\phi'
		+\frac{Q_e^2}{r^3N f_i(\phi)}
		\left(
		\phi'-\frac{\dot f_i(\phi)}{2 r f_i(\phi) }
		\right)=0\ .
 	\end{equation}		
	}.

	To solve the set of ordinary differential equations (ODEs) \eqref{E24}-\eqref{E28}, 
	we have to implement suitable boundary conditions for the desired functions $\big( m,\ \delta,\ \phi,\ V \big) $ and corresponding derivatives. Near the BH event horizon, located at $r=r_H >0$\ze ,
	the  solutions possess a power series expansion 
	\begin{eqnarray}
	 && m\ze (r) = \frac{r_H}{2}+m_1 (r-r_H) + \ze 	\dots \ ,\qquad \qquad
	  \delta\ze (r) = \delta _0 + \delta _1 (r-r_H)+ \dots \ , \nonumber\\
	  && \phi \ze (r) = \phi _0 +\phi _1 (r-r_H)+ \ze \dots\ ,\qquad \qquad \ \ 
	  V\ze (r) =  v_1 (r-r_H) + \ze \dots \ , \label{E29}
	\end{eqnarray}
	where
	\begin{equation}\label{E210}
 	 m_1 = \frac{ Q_e ^{\ze 2}}{2\ze\ze f_i (\phi _0)\ze\ze r_H^{\ze 2}}\ , \qquad
	\phi_1=\frac{\dot f_i(\phi_0)}{2r_H f_i(\phi_0)}\frac{Q_e^2}{Q_e^2-r_H^2 f_i(\phi_0)} \ , 
\qquad
	 \delta _1 = -\phi _1 ^{\ze \ze 2}\ze \ze r_H\ , \qquad 
	 v_1 = -\frac{e^{-\delta _0} Q_e}{r_H^ 2 f_i \ze (\phi _0)}\ ,
	\end{equation}
in terms of two essential parameters
 $\phi _0$ and $\delta _0$.
		The horizon data fixes the values of the Hawking temperature
 $T_H = \frac{1}{4\pi}N' (r_H)\ze e^{-\delta (r_H)}$, 
 and horizon area, $A_H =4\pi\ze  r_H ^{\ze\ze 2}$.
The expression of the Kretschmann scalar, $K\equiv R_{\mu\nu\alpha\beta} R^{\mu\nu\alpha\beta}$,  and the energy 
density $\rho=-T_t^t$ at the horizon are also
of interest
\begin{eqnarray}
K(r_H)=\frac{4}{r_H^4}
\left[
3-\frac{6 Q_e^2}{r_H^2 f_i(\phi_0)}+\frac{5 Q_e^4}{r_H^4 f_i^2(\phi_0)}
\right ]\ , \qquad
\rho(r_H)=\frac{Q_e^2}{2r_H^4 f_i(\phi_0)} \ ,
\label{ed}
\end{eqnarray}
while the Ricci scalar vanishes as $r\to r_H$. For future reference observe the energy density
$\rho(r_H)$
 vanishes when the coupling blows up and changes sign when the coupling changes sign.

An asymptotic approximation of the solution in the far field takes the form:
	\begin{eqnarray}
	 && m(r) = M-\frac{Q_e ^{\ze \ze 2}+Q_s ^{\ze\ze 2}}{2\ze\ze r}+\ze \dots \ ,\qquad \qquad
	 \phi (r) = \frac{Q_s}{r}+\frac{Q_s\ze M}{r^2}+\ze \dots \ , \nonumber\\
	 && V(r) = \Phi +\frac{Q_e}{r}+\ze \dots \ , \qquad \qquad \qquad \quad \ 
	 \delta (r) = \frac{Q_s ^{\ze\ze 2}}{2\ze\ze r^2}+\ze \dots \ . \label{E211}
	\end{eqnarray}
	This expansion introduces another three constants: the ADM mass $M$, 
	the electrostatic potential at infinity $\Phi$ and the scalar charge $Q_s$. 
	The full equations of motion can now be integrated with these asymptotic behaviours.

The solutions satisfy 
	the \textit{virial identity}~\cite{Herdeiro:2018wub}, 

	\begin{equation}\label{E212}
	 \int _{r_H} ^{\infty} dr \left\{ e^{-\delta}\ze r^2 \ze \phi ^{' 2} \left[1+\frac{2\ze r_H}{r}\Big(\frac{m}{r}-1 \Big)\right]\right\} =  \int _{r_H} ^{\infty} dr \left[\frac{e^{-\delta}}{f_i \ze (\phi )}\left( 1-\frac{2\ze r_H}{r}\right) \frac{Q_e ^{\ze 2}}{r^2}\right] \ ,
	\end{equation}
which is obtained via a scaling argument, 
and the Smarr relation~\cite{Bardeen:1973gs,smarr1973mass}, 
which turns out not to be affected by the scalar hair~\cite{Herdeiro:2018wub}, 
	\begin{equation}\label{E213}
	 M=\frac{1}{2}\ze T_H A_H + \Phi   Q_e\ .
	\end{equation}
 The first law of BH thermodynamics is $dM = \frac{1}{4}T_H\ze dA_H +\Phi dQ_e$. 
The solutions satisfy also the following relation~\cite{Herdeiro:2018wub}
	\begin{equation}
	\label{E214}
	 M^2 + Q_s ^{\ze\ze 2}= Q_e ^{\ze \ze 2} + \frac{1}{4}T_H^2 A_H^2\ .
	\end{equation}
	Remarkably, one can show that \eqref{E214}, dubbed \textit{non-linear Smarr} relation, holds for any $f_i \ze (\phi )$ 
	that behaves as $\phi \rightarrow Q_s /r$ asymptotically ($i.e.$ as $r\rightarrow \infty $). 
	
%
%%%%%%%%%%%%%%%%%%%%%%%%%%%%%%%%%%%%%%%%%%%
%%%%%%%%%%%%%%%%%%%%%%%%%%%%%%%%%%%%%%%%%%%
	\subsection{The coupling functions}\label{S21}
%%%%%%%%%%%%%%%%%%%%%%%%%%%%%%%%%%%%%%%%%%%
%%%%%%%%%%%%%%%%%%%%%%%%%%%%%%%%%%%%%%%%%%%
%
	The coupling function $f_i (\phi ) $ must obey the following criteria: 1) accommodate non-scalarised solutions, which amounts to the condition 
	$\dot{f}_i(0)=0$. This can be intrepreted as implementing a $\mathbb{Z}_2$ symmetry $\phi\rightarrow -\phi$; 2) the form of the coupling is  constrained by two Bekenstein type identities~\cite{bekenstein1972transcendence}, which require
		\begin{equation}\label{E215}
		\ddot{f} _i > 0\ , \qquad \qquad \phi\ze\ze \dot{f} _i  > 0\ ,\\ 
		\end{equation}	 
for some range of the radial coordinate;
3) obey $f_i(0)=1$, so that one recovers Maxwell's theory near spatial infinity. 
	In this work we will consider four forms for the coupling constant consistent with the above requirements: 
		\begin{itemize}
		 \item[i)] an exponential coupling, $f _E (\phi) = e^{-\alpha\ze \phi ^2}$, first used in this context in~\cite{Herdeiro:2018wub}; 
		  \item[ii)] a hyperbolic cossine coupling, $f _C (\phi) = \cosh({\sqrt{2|\alpha|} \phi}$); 
		 \item[iii)] a power coupling, $f _P (\phi) = 1-\alpha\ze \phi ^2$, already discussed in this context in~\cite{Boskovic:2018lkj};
		 \item[iv)] a fractional coupling, $f _F (\phi) = \frac{1}{1+\alpha\ze \phi ^2 }$.
		\end{itemize}
	 The coupling constant $\alpha$ is a dimensionless constant in all cases, and, except for the hyperbolic function, the conditions on $f_i$ imply that $\alpha <0$ for a purely electric field, $i.e.$ $F_{\mu\nu}F^{\mu\nu} <0$. The  $f_i$ candidates shall be specified by the subscript $i \in \{ E, \ C,\ P,\ F\}$, respectively. For $|\alpha| \ze \phi ^2 \ll 1\ze $ (and $\alpha<0$), $f_E$, $f_C$ and $f_F$ possess the same Taylor expansion to first order which coincides with the (exact) form of $f_P$:
		\begin{equation}
		 f_F \ze (\phi ) \approx  f_C \ze (\phi ) \approx f_E \ze (\phi ) \approx 1+|\alpha|\ze\ze \phi ^2 +\mathcal{O}\ze (\phi ^4)\ .
		\label{E216}
		\end{equation}	 
	This observation implies, in particular, that the zero mode coincides for all cases in the spherical sector, fundamental branch, scalarised solutions. Thus, from~\cite{Herdeiro:2018wub}, scalarised solutions exist in all  cases for $\alpha <-1/4$.
	Fixing $-\alpha>1/4$ scalarised solutions exist above a certain threshold for the charge to mass ratio $q$. From another perspective,  there is minimum value of $|\alpha|$ for each $q$ of a RN BH in order for scalarised solutions to exist. This minimum value corresponds to the branching point and is presented in Table~\ref{T0}  for some values of $q$. 
	As the scalar field increases and non-linearities become relevant, the differences between the models with different couplings emerge.

		\begin{table}[h!]
 	 	 \centering
 	 	 \caption{Minimum value of $|\alpha|$ for scalarisation of a RN BH with charge to mass ratio $q$.}
 	 	 \vspace{2mm}
 	 		\begin{tabular}{||c||c|c|c|c|c|c|c|c|c|c||}
			\hline\hline
                         $q$ & $1.0$  & $0.9$  & $0.8$  & $0.7$ & $0.6$  & $0.5$  & $0.4$ & $0.3$ & $0.2$ & $0.1$\\ 	 
			 \hline
			 $|\alpha|$  & $0.25$ & $2.995$ & $5.121$  & $8.019$ & $12.37$  & $19.50$ & $32.56$ & $60.72$ & $141.0$ & $574.9$ 
			 \label{T0}
			\end{tabular}
 		 \end{table}

	%
%%%%%%%%%%%%%%%%%%%%%%%%%%%%%%%%%%%%%%%%%%%%%%%%%%%%%%%%%%%%%%%%%%%%%%%%%%%%%%
%%%%%%%%%%%%%%%%%%%%%%%%%%%%%%%%%%%%%%%%%%%%%%%%%%%%%%%%%%%%%%%%%%%%%%%%%%%%%%
\section{Numerical results}\label{S3}
%%%%%%%%%%%%%%%%%%%%%%%%%%%%%%%%%%%%%%%%%%%%%%%%%%%%%%%%%%%%%%%%%%%%%%%%%%%%%%
%%%%%%%%%%%%%%%%%%%%%%%%%%%%%%%%%%%%%%%%%%%%%%%%%%%%%%%%%%%%%%%%%%%%%%%%%%%%%%
%

		The set of four ODEs~\eqref{E24}-\eqref{E27} can be numerically solved through a Runge-Kutta strategy, given the aforementioned boundary conditions. Our numerical method implements a six(five) Runge-Kutta integration algorithm (RK65) with an adaptative step size and a shooting method. The latter is implemented in the unknown parameters and ensures the fulfillment of the boundary conditions. This code is written in \textsc{c} and was developed and extensively tested by us. 
%
%%%%%%%%%%%%%%%%%%%%%%%%%%%%%%%%%%%%%%%%%%%
%%%%%%%%%%%%%%%%%%%%%%%%%%%%%%%%%%%%%%%%%%%
	\subsection{Solutions profile}\label{S31}
%%%%%%%%%%%%%%%%%%%%%%%%%%%%%%%%%%%%%%%%%%%
%%%%%%%%%%%%%%%%%%%%%%%%%%%%%%%%%%%%%%%%%%%

\begin{figure}[h]
		  \centering
	 	 \includegraphics[scale=0.4]{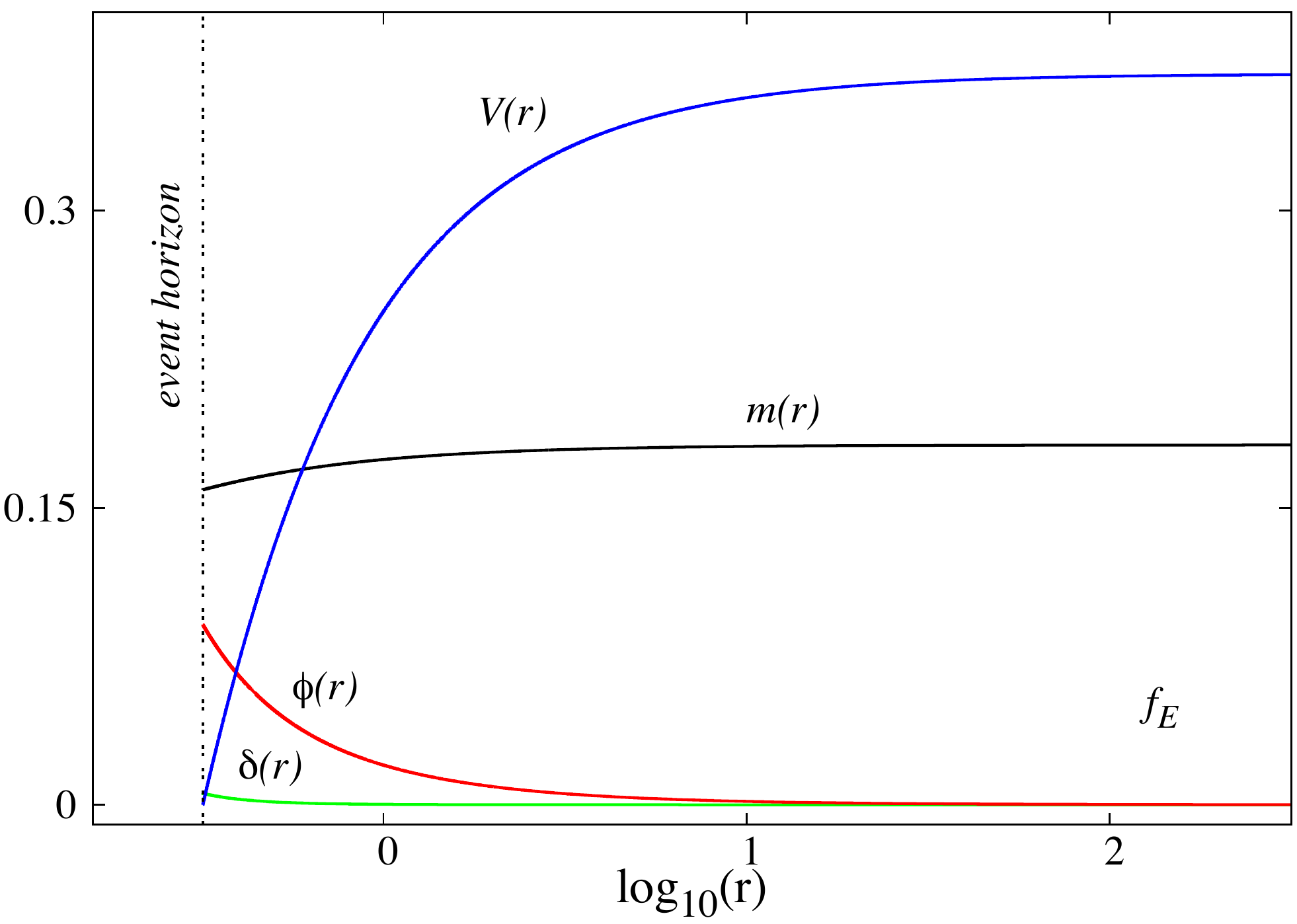}\hfill
	 	 \includegraphics[scale=0.4]{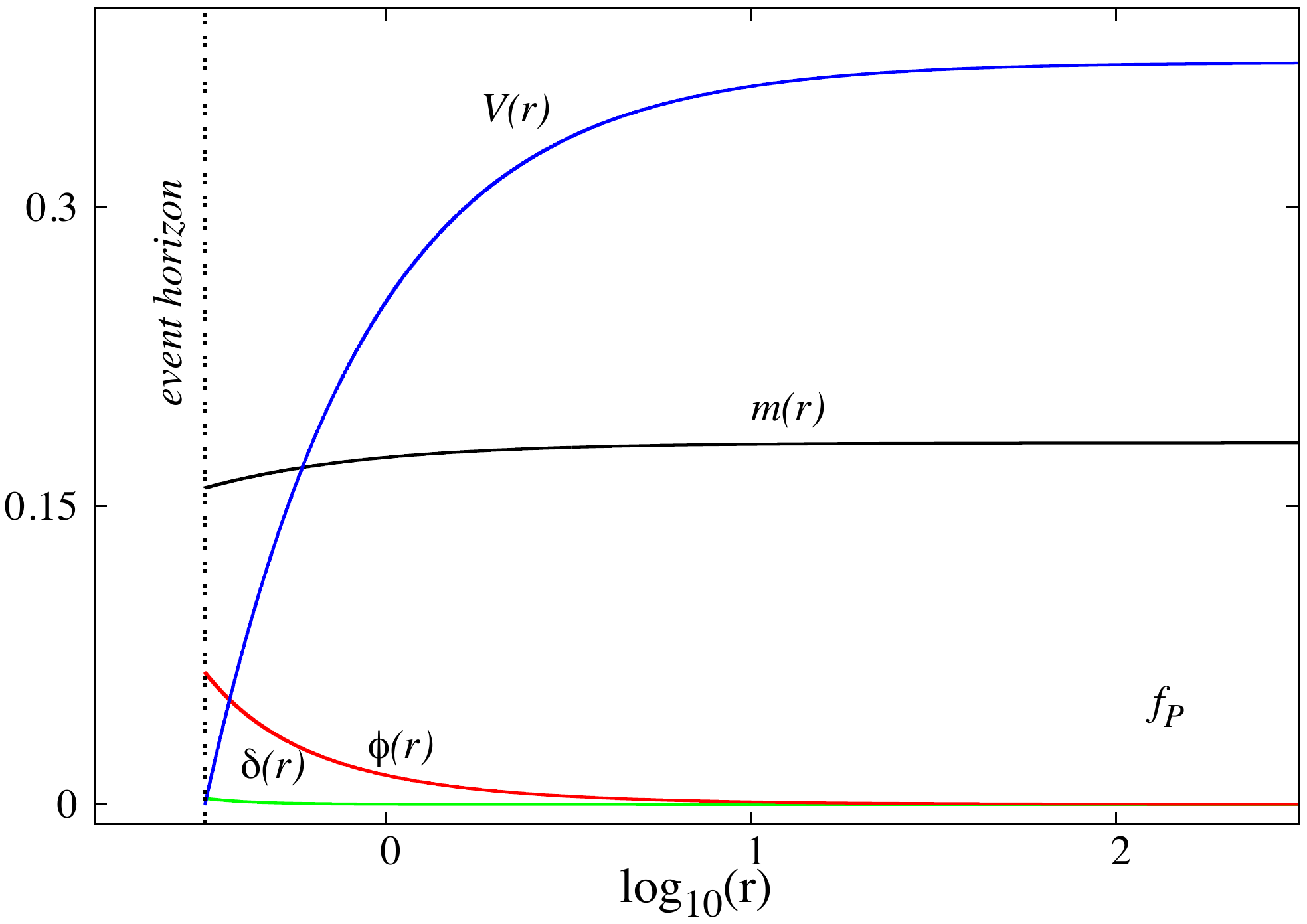}\\
	 	 \includegraphics[scale=0.4]{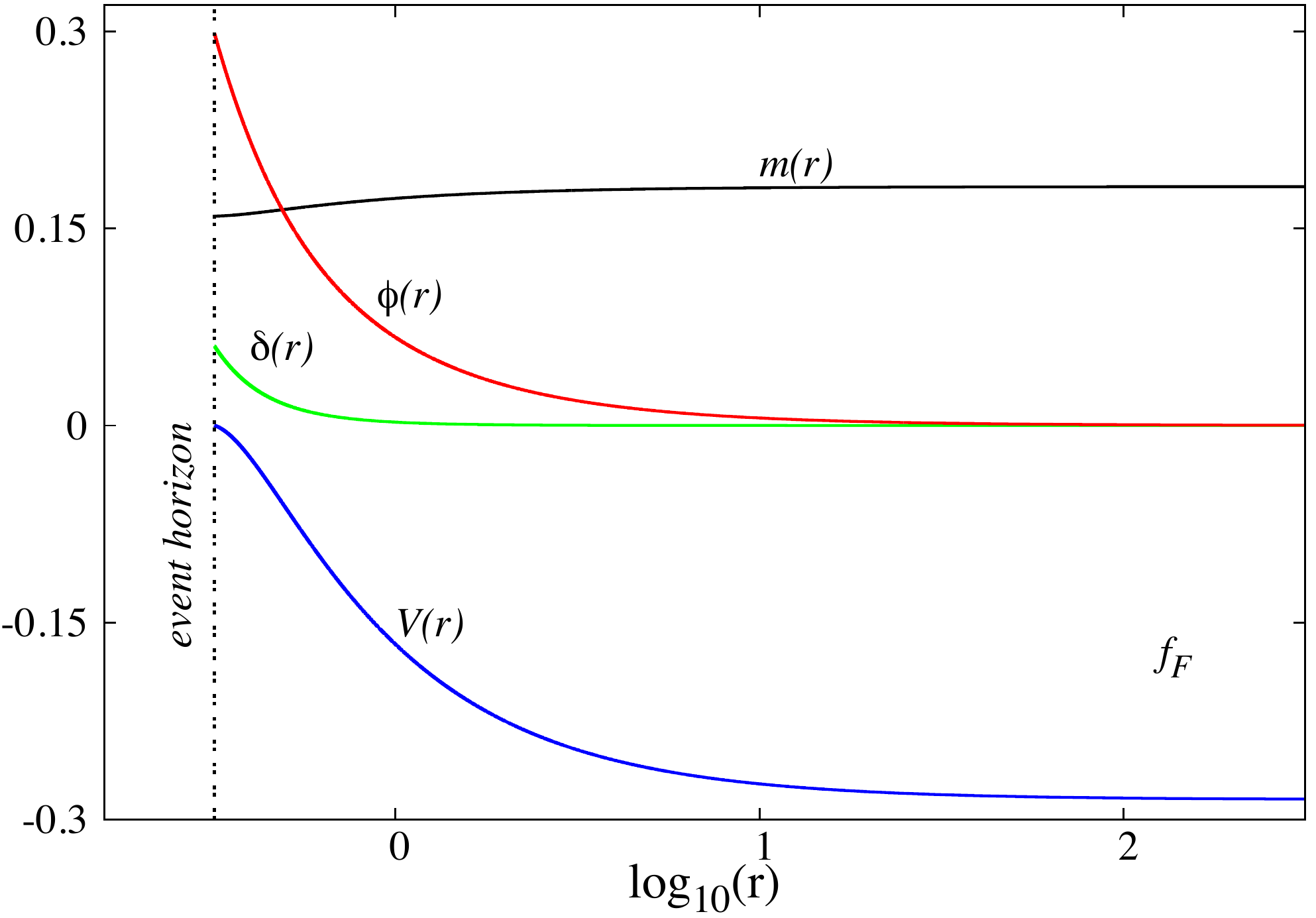}
	 	 \caption{Scalarised BH radial functions for $\alpha =-10$ and $q=0.66$. (Top left panel) $f_E$; (top right panel) $f_P$; (bottom panel) $f_F$.}
	 	 \label{F2}
		\end{figure}

Let us start by exhibiting some typical solutions obtained from the numerical integration. 	In Fig.~\ref{F2} the various radial functions defining the scalarised BHs are represented for an illustrative coupling of $\alpha=-10$, charge to mass ratio $q\equiv Q/M=0.66$ and for three different choices of coupling. A universal feature of those nodeless solutions is that the scalar field is monotonically decreasing function of the radius. 
Thus the scalar field value at the horizon, $\phi_0$, $cf.$~\eqref{E29}, is always the maximum of the scalar field. The scalar field vanishes asymptotically, $cf.$~\eqref{E211}. In fact, at far enough radius ($r>10 ^2$), all defining functions of  the scalarised BHs converge to the ones of a comparable ($i.e.$ with the same global charges) RN BH.  Another typical feature illustrated by the figure is that the differences between the exponential and power-law couplings are small -- see Table~\ref{T1} (and the same would apply to the $\cosh$ coupling, thus not shown), and more pronounced for the fractional coupling. Yet, for the same values of $\alpha$ and $q$ the scalarisation in the exponential coupling is stronger than for the power law one (and intermediate in the $\cosh$ one); this is visible in the value of the scalar field at the horizon on the two top panels of the figure. We remark that these data are well within the numerical errors: our tests have exhibited a relative difference of $10 ^{-8}$ for the virial relation; $10^{-7}$ for the Smarr relation and $10^{-6}$ to the non-linear Smarr relation.

		\begin{table}[h!]
 	 	 \centering
 	 	 \caption{Characteristic quantities for  scalarised BH solutions with four choices of couplings,  $\alpha =-10$ and $q=0.66\ze $. $a_H$ is the reduced horizon area, $a_H\equiv A_H/16\pi M^2$.}
 	 	 \vspace{2mm}
 	 		\begin{tabular}{c|cccccc}
 	  		 $f_i (\phi )$  & $r_H$ & $M$ & $Q_s$ & $\Phi$ & $a_H$ & $T_H$\\
 	  			
			 \hline
			 $f_E$ & $0.3180$ & $0.1816$ & $0.0167$ & $0.3689$ & $0.7663$ & $0.2162$ \\
			 
			 $f_C$ & $0.3180$ & $0.1816$ & $0.0132$ & $0.3720$ & $0.7663$ & $0.2156$ \\
		
			 $f_P$ & $0.3180$ & $0.1816$ & $0.0122$ & $0.3729$ & $0.7663$ & $0.2154$ \\
			 	 
			  $f_F$ & $0.3186$ & $0.1818$ & $0.0561$ & $0.2848$ & $0.7680$ & $0.2314$ 
			 \label{T1}
			\end{tabular}
 		 \end{table}

    	For the particular case of the fractional coupling, however, a different type of solutions, that we call \textit{exotic}\ is possible. If $1+\alpha \phi_0^2<0$, then the corresponding solutions have a region of negative energy density in the vicinity of the horizon, $cf.$~\eqref{ed} and Fig.~\ref{F1} (right panel). Moving away from the horizon, as the value of the scalar field decreases monotonically, $cf.$ Fig.~\ref{F1} (left panel), it passes through the point at which the coupling diverges. This divergence is, however, benign and the geometry is smooth therein.  {
			This can be understood from the equations~\eqref{s1}, which contain $1/f_F$ terms but no divergencies. 
			Moreover, beyond a critical radius the energy density is again positive - Fig.~\ref{F1} (right panel inset) . The  negative energy region in the vicinity of the horizon leads to a decrease in the mass function profile - see Fig.~\ref{F1} (left panel).

		\begin{figure}[h]
		 \centering
	 \includegraphics[scale=0.4]{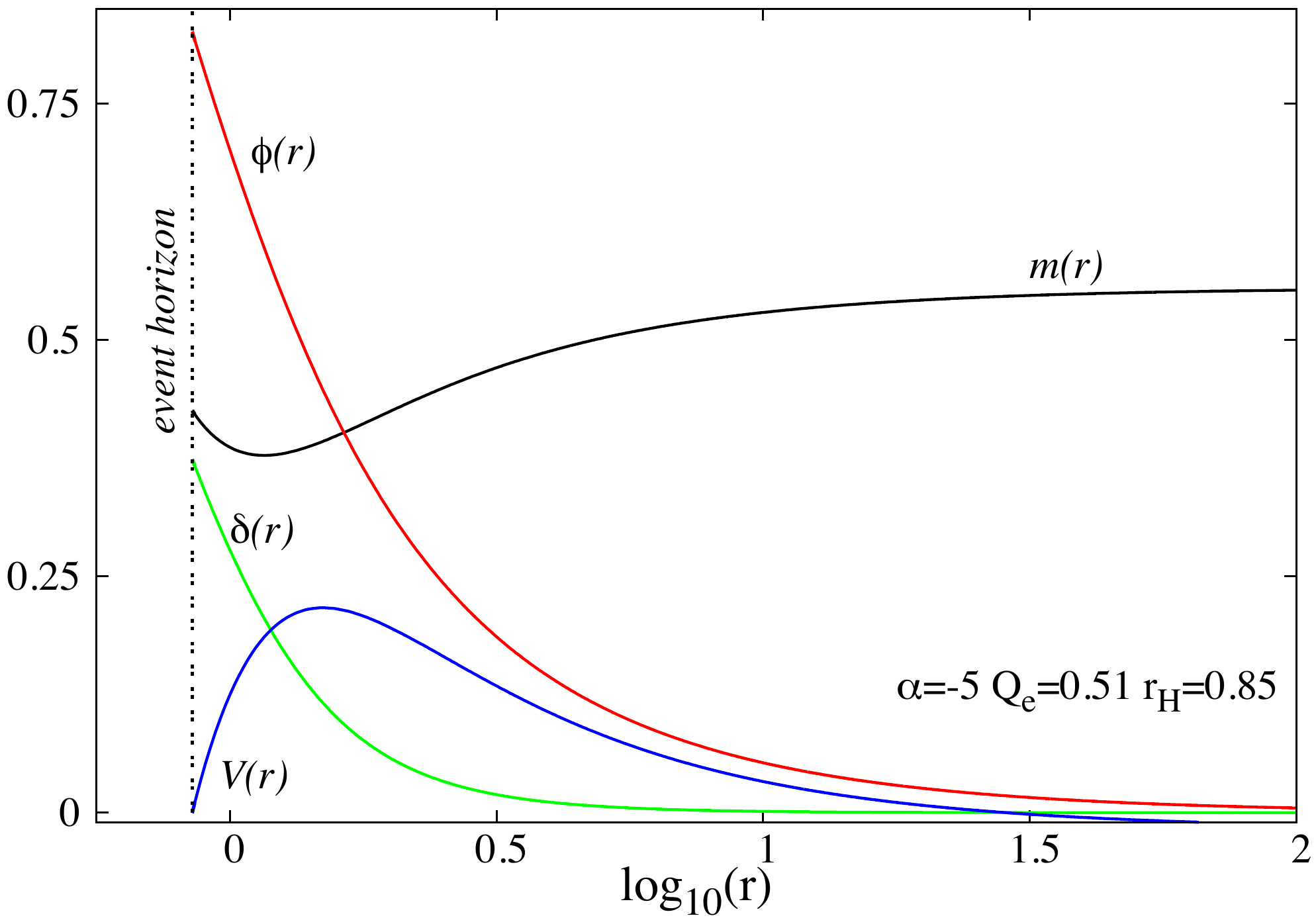}\hfill
 \includegraphics[scale=0.4]{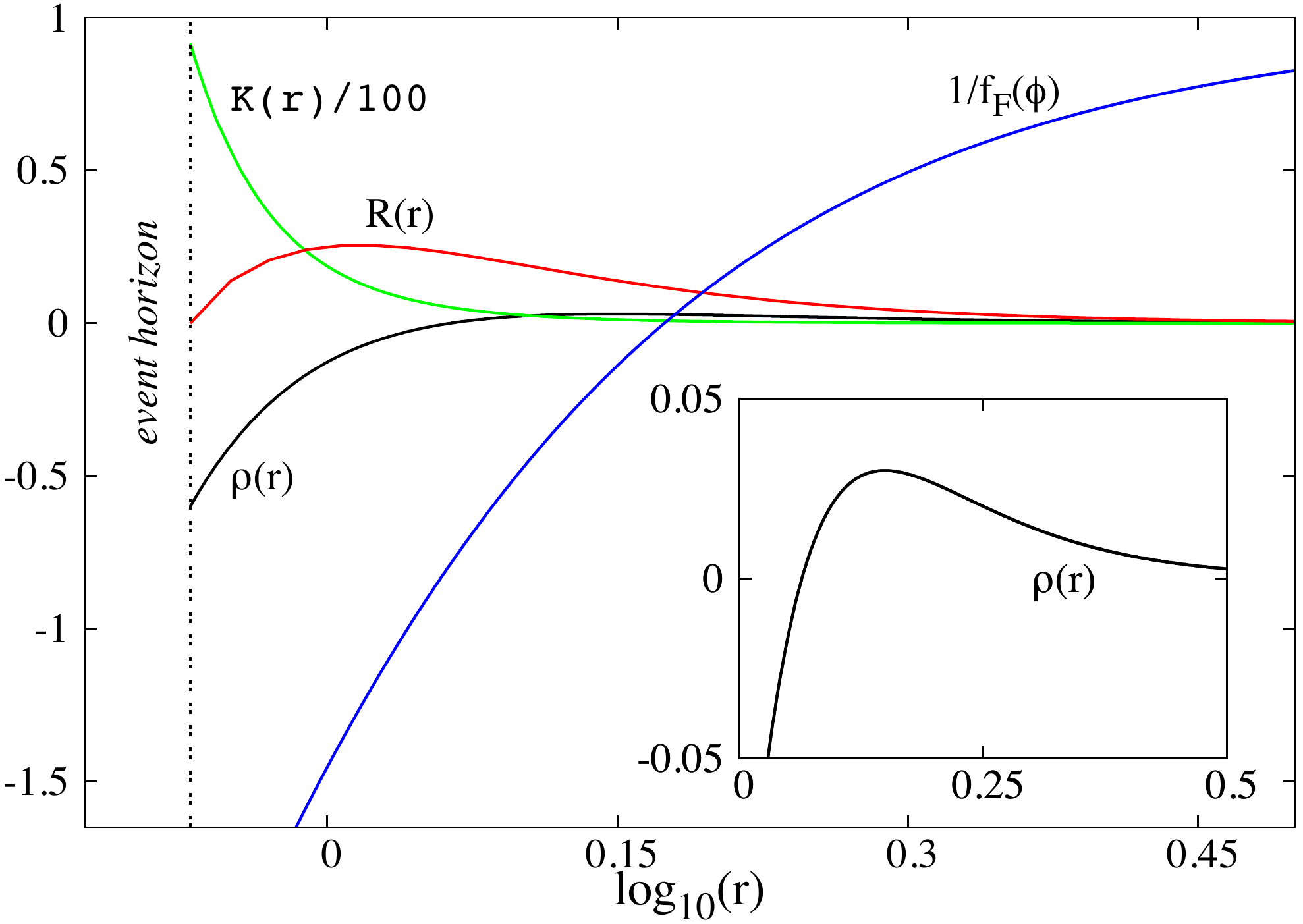}\hfill
	 	 \caption{A typical scalarised BH in an EMS model with the coupling function $f_F$, 
	which possesses a region with negative energy density, $\rho<0$.
		(Left panel) Profiles of the metric and matter functions; (right
		panel) the energy density (zoom in presented in the inset), the Ricci and Kretschmann scalars and the inverse of the coupling function $f_F(\phi)$
		which  changes sign at some finite $r$. 
		This plot manifests that  solutions with $\rho<0$ are smooth.
		}
		\label{F1}
		\end{figure}

%
%%%%%%%%%%%%%%%%%%%%%%%%%%%%%%%%%%%%%%%%%%%
%%%%%%%%%%%%%%%%%%%%%%%%%%%%%%%%%%%%%%%%%%%
	\subsection{Domain of existence}\label{S32}
%%%%%%%%%%%%%%%%%%%%%%%%%%%%%%%%%%%%%%%%%%%
%%%%%%%%%%%%%%%%%%%%%%%%%%%%%%%%%%%%%%%%%%%
%
Let us now focus on a comparative study of the domain of existence for the scalarised, fundamental, spherically symmetric solutions for the  chosen couplings.	Action \eqref{E21} imposes the scalar field equation of motion
		\begin{equation}\label{E318}
		 \Box \phi =\frac{1}{4}\dot  f_i (\phi ) \mathcal{I}\ ,
		\end{equation}
	which, after being linearized around a scalar free solution, yields 
	$\big(\Box -\mu _{\rm eff} ^{\ze\ze 2} \big)\delta \phi = 0\ze $, 
	where
	$\mu _{\rm eff} ^{\ze\ze 2} = - \ddot{f}_i (0) \mathcal{I}/4\ze 
	= -|\alpha|\ze\ze Q_e^2 \ze r^{-4}  $.
	In order for a tachyonic instability to settle in, we must have $ \mu _{\rm eff} ^{\ze\ze 2} <0\ze $. 
	The spherical symmetry allows a scalar field's decomposition in (real) spherical harmonics,
	$\delta \phi (r,\theta, \phi )=\sum _{\ell\textbf{m}} Y_{\ell\ze\textbf{m}} (\theta,\phi )\ze U_{\ze \ell} (r)$. 
	The scalar field equation simplifies to
		\begin{equation}
		\label{E319}
	 	 \frac{e^\delta }{r^2} \left( \frac{r^2 N  }{e^\delta }\ze U' _{\ze \ell} \right) '-\left[\frac{\ell(\ell+1)}{r^2}+\mu _{\rm eff} ^{\ze \ze 2} \right] U_\ell=0\ ,
		\end{equation}
which is an eigenvalue problem: fixing the coupling $\alpha$, for a given $\ell$, requiring an asymptotically vanishing, smooth scalar field, selects a discrete set of BHs solutions, $i.e.$ RN solutions with a certain $q$. These are the bifurcation points of the scalar-free solution. They are labelled by an integer $n\in \mathbb{N}_0$; $n=0$ is the fundamental mode, whereas $n>1$ are excited states (overtones). One expects only the fundamental solutions to be stable~\cite{Myung:2018jvi}. Focusing on the latter, solutions with a smaller (larger) $q$ are stable (unstable) against spherical scalar perturbations, for that coupling. 
			Clearly, for any $f_i\ze (\phi )$, setting $\delta =0$ and $N(r) =1-2M/r+Q_e^2 / r^2$ in \eqref{E21} allows us to recover the usual RN metric. Then, a scalarised solution can be dynamically induced by a scalar perturbation of the background, as long as the scalar-free RN solution is in the unstable regime.

As pointed out in~\cite{Herdeiro:2018wub}, for $\ell=0$, one finds the following exact solution\footnote{No exact solution appears to exist for $\ell\geqslant 1$, and equation (\ref{E319}) is solved numerically. These modes, nonetheless, 
also possess non-linear continuations leading to static, non-spherically symmetric scalarized BHs \cite{Herdeiro:2018wub}.}
\begin{eqnarray}
\label{ex1}
U(r)=P_u 
\left[
1+\frac{2Q_e^2(r-r_H)}{r(r_H^2-Q_e^2)}
\right]\ , \qquad {\rm where} \qquad u\equiv\frac{1}{2}(\sqrt{4\alpha+1}-1) \ ,
\end{eqnarray}
$P_u$ being a Legendre function.
For generic
parameters
 $(\alpha,Q_e,r_H)$, 
the function
$U(r)$ 
approaches a constant \textit{non-zero} value as $r\to \infty$,
\begin{eqnarray}
\label{ex2}
U(r) \to 
{}_2F_1 
\left[
\frac{1}{2}(1-\sqrt{4\alpha+1}),\frac{1}{2}(1+\sqrt{4\alpha+1}),1; \frac{x^2}{x^2-1}
\right]+\mathcal{O}\left(\frac{1}{r}\right) \ ,
\end{eqnarray}
where $x=Q_e/r_H$.
Thus finding the $\ell=0$ unstable mode
of the RN BH reduces to a study of the zeros of the hypergeometric function
${}_2F_1 $. Some values were given in Table~\ref{T0}.

The solution  of \eqref{E319} yields a RN BH surrounded by a vanishingly small scalar field. The
full set of such configurations make up the \textit{existence line} which, as discussed before, is \textit{common} for all specific coupling functions discussed herein, as they are identical for small $\phi$. The differences in the domain of existence of the four couplings emerge for larger values of $\phi$, wherein non-linearities become important.
    
    The domains of existence for the scalarised BHs with the $f_E$, $f_C$, $f_P$ couplings are exhibited in Fig.~\ref{F3} (left panel). They are delimited by the \textit{existence line} -- (dashed blue) on which the RN BHs that support the zero mode exist -- and a \textit{critical line} -- (solid red) which corresponds to a singular scalarised BH configuration. In between (shaded blue regions: dark for $f_P$, dark+medium for $f_C$, dark+medium+light for $f_E$), scalarised BHs exist. In particular, for $q=Q_e/M \leqslant 1$ the usual RN BH and the scalarised  solutions co-exist with the same global charges.   In this region there is non-uniqueness.  The scalarised solutions are always entropically favoured (see Section~\ref{S4}). These spherical scalarised BHs are candidate endpoints of the spherical evolution (if adiabatic) of the linearly unstable RN BHs in the EMS model.
    
		\begin{figure}[h]
		 \centering
	 	 \includegraphics[scale=0.4]{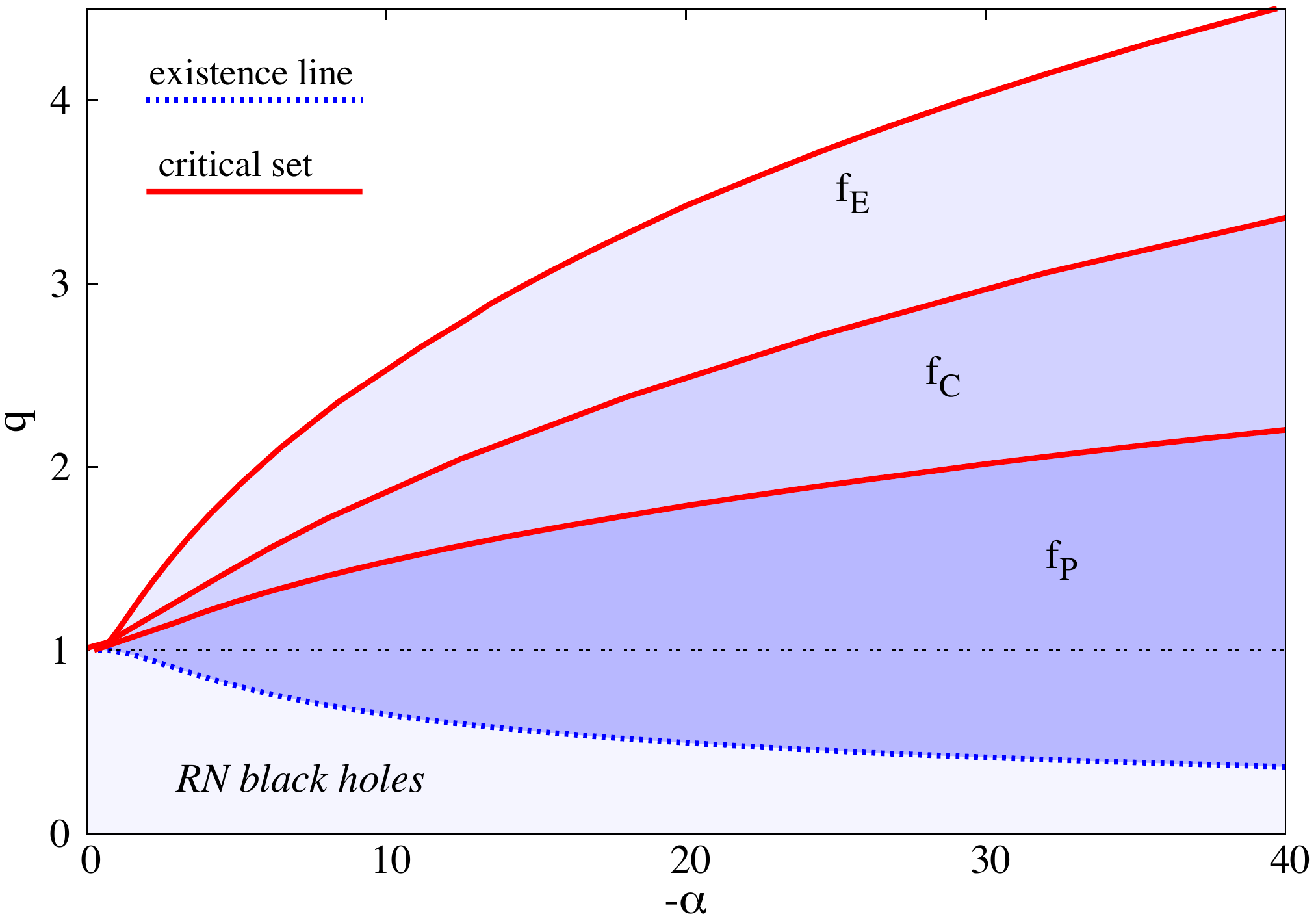}\hfill
	 	 \includegraphics[scale=0.4]{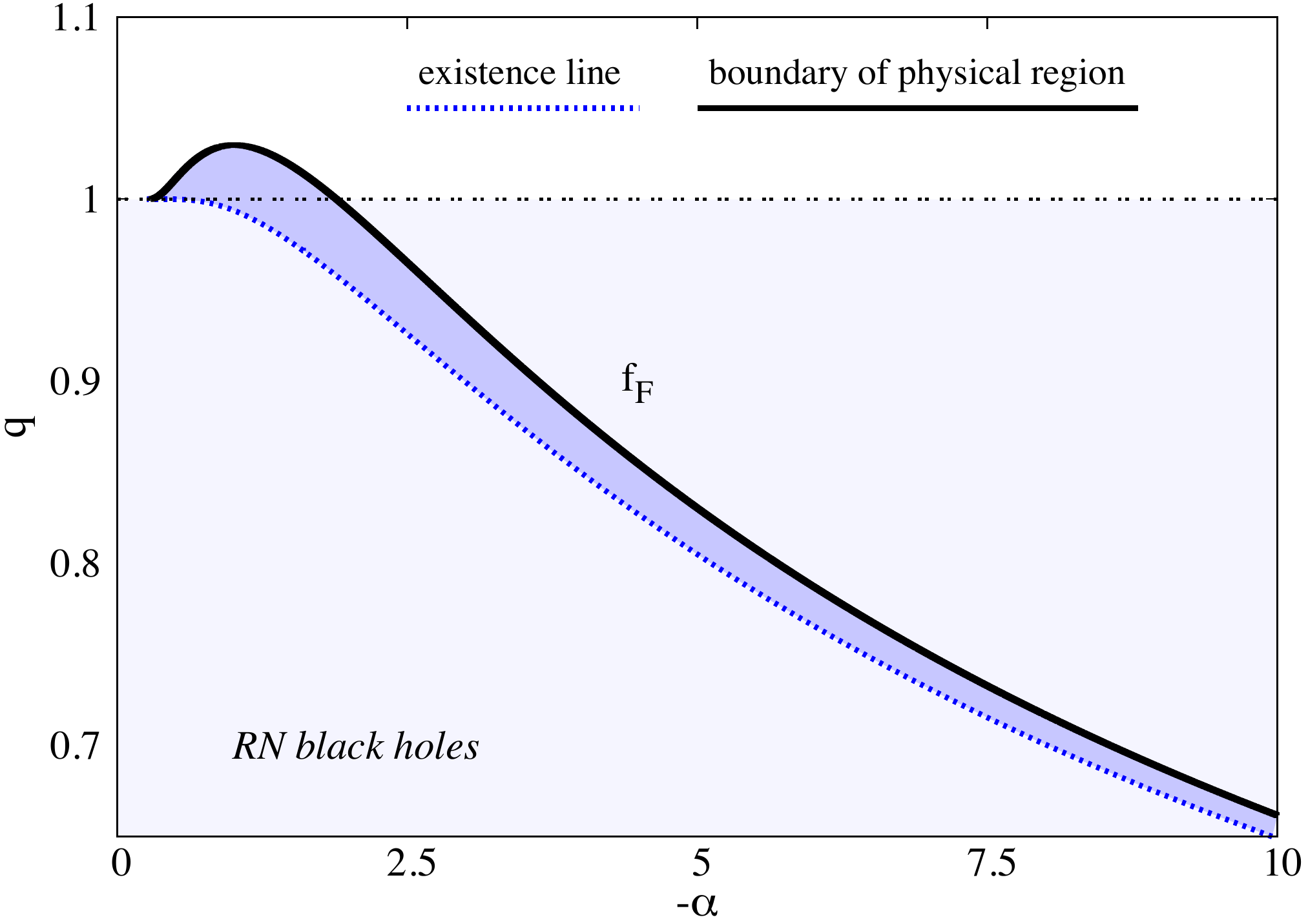}
	 	 \caption{Domain of existence of scalarised BHs in EMS models (shaded blue regions). The domain of existence is always delimited by the existence line (dashed blue line) and the critical (red) line. (Left panel) $f_E\ze (\phi )$, $f_C(\phi)$ and $f_P\ze (\phi )$ couplings. (Right panel) $f_F\ze (\phi )$ coupling. Here we only exhibit the physical region, which is delimited by the existence line and the line at which the coupling function diverges at the horizon. The latter is the boundary of the physical region; above it, solutions have a negative energy density in the vicinity of the horizon.}
	 	 \label{F3}
	\end{figure}

	At the critical line, numerics suggest $K\rightarrow \infty$, $T_H,A_H\rightarrow 0$, while $M,Q_s$ remain finite. As another feature, along  $\alpha =$constant branches, $q$ increases beyond unity: therefore, scalarised BHs can be overcharged~\cite{Herdeiro:2018wub}.

Comparing the domain of existence of the exponential, $\cosh$ and power-law couplings  (Fig.~\ref{F3}, left panel) we see that they are qualitatively similar. The critical set for the same $\alpha$, however,  occurs at the smallest value of $q$ for the power law coupling, an intermediate value for the hyperbolic coupling and the largest value of $q$ for the exponential coupling. So, the exponential coupling allows \textit{maximising} the possibility of overcharging the BH and, in this sense, of maximising the differences with the RN BH case. Moreover, as seen before $cf.$ Fig.~\ref{F2}, scalarisation is ``stronger" for the $f_E$ coupling than for $f_P$ (with an intermediate value for $f_C$). We also remark that for a given $\alpha $, as $q$ increases, so does the scalar field's initial amplitude $\phi _0 \ze $. As already mentioned, the scalar field profile is always such that the scalar field is monotonically decreasing.  Thus, the global maximum of the scalar field occurs at the BH horizon, and increases, for fixed $\alpha$, with $q$, and one can take $\phi_0$ as a measure of $q$ and vice-versa.

The domain of existence of the $f_F$ coupling function (Fig.~\ref{F3} - right panel) can be divided into two parts. For $\alpha =$constant,  $\phi_0$ grows from the existence line until it reaches $\phi_0^2=1/|\alpha|$ at the \textit{divergence line}, corresponding to the pole of the coupling. These solutions span the physical region wherein solutions have a positive energy density.  Beyond the divergence line solutions have $\phi_0^2>1/|\alpha|$ and thus a negative energy density region near the horizon extending  up to a critical radius 
at which $\rho=0$ - see Fig.~\ref{F1}.
Beyond this point the energy density is again positive. Solutions in the exotic region appear to be smooth exhibiting no other obvious pathologies apart from the negative energy density. The physical region of the domain of existence will tend to thin down to zero, as $|\alpha|$ increases. 
   Unlike the other studied couplings, for a model with $f_F$, the scalarised BH can only be overcharged and in the physical region if the coupling constant is in a compact interval: $\alpha \in [\ze -1.89074,\ze -1/4\ze]$, with a maximum of $q=1.02971$ for $\alpha =-1.0115\ze $ - $cf.$ Fig.~\ref{F3} - right panel.

 %%%%%%%%%%%%%%%%%%%%%%%%%%%%%%%%%%%%%%%%%%%
%%%%%%%%%%%%%%%%%%%%%%%%%%%%%%%%%%%%%%%%%%%
	\subsection{Perturbative stability}\label{S33}
%%%%%%%%%%%%%%%%%%%%%%%%%%%%%%%%%%%%%%%%%%%
%%%%%%%%%%%%%%%%%%%%%%%%%%%%%%%%%%%%%%%%%%%
Following a standard technique for studying perturbative stability against radial perturbations, we consider spherically symmetric, linear perturbations of our equilibrium solutions,
keeping the metric ansatz (\ref{E22}),
 but allowing the functions $N$, $\delta$ and $\phi,V$ to depend on $t$ as well as on $r$:
\begin{eqnarray}
\label{stab1}
ds^2=- \tilde N(r,t)e^{-2 \tilde \delta(r,t)} dt^2+\frac{dr^2}{\tilde N(r,t)}+r^2(d\theta^2+\sin^2 \theta d\varphi^2) \ ,
 \qquad A= \tilde V(r,t) dt \ , \qquad \phi=\tilde \phi(r,t) \ .
\end{eqnarray}
The time dependence enters as a periodic perturbation with frequency $\Omega$, for each of these functions:
\begin{eqnarray}
\label{stab3}
&&
 \tilde N(r,t)=N(r)+\epsilon N_1(r)e^{-i \Omega t}\ , \qquad  \tilde \delta(r,t)=\delta(r)+\epsilon \delta_1(r)e^{-i \Omega t} \ , 
\\
&&
\nonumber
 \tilde \phi(r,t)=\phi(r)+\epsilon \phi_1(r)e^{-i \Omega t}\ , \qquad \tilde V(r,t)=V(r)+\epsilon V_1(r)e^{-i \Omega t}\ .
\end{eqnarray}
From the linearised field equations around the background solution, the metric perturbations and $V_1(r)$ can be expressed in terms of the scalar field perturbation,
\begin{eqnarray}
\label{stab4}
N_1=-2r N\phi' \phi_1 \ , \qquad \delta_1=-2\int dr ~r\phi' \phi_1' \ , \qquad V' _1= -V'\left[ \delta_1  + \phi_1 \frac{\dot f_i(\phi)}{f_i (\phi)}\right] \ ,
\end{eqnarray}
thus yielding a single perturbation equation for $\phi_1$.
This equation can be written in  the standard Schr\"odinger-like form:
\begin{eqnarray}
\label{stab5}
-\frac{d^2 }{dx^2}\Psi+U_{\Omega} \Psi=\Omega^2 \Psi \ ,
\end{eqnarray}
where we have defined $\Psi\equiv r \phi_1$ and the `tortoise' coordinate $x$ by
\begin{eqnarray}
\label{stab6}
\frac{dx}{dr}=\frac{1}{e^{-\delta} N}\ . \qquad  
\end{eqnarray}
The perturbation potential $U_{\Omega}$ is defined as:
\begin{equation}
\label{stab7}
U_{\Omega}\equiv \frac{e^{-2\delta}N}{r^2}
\left\{ 
1-N-2r^2 \phi'^2
-\frac{Q_e^2}{2r^2}
\left[
\frac{2}{f_i(\phi)}(1-2r^2 \phi'^2)
-\frac{2\dot f_i^2(\phi)}{f_i^3(\phi)}
+\frac{1}{f_i^2(\phi)}
(\ddot f_i(\phi)+4r\phi' \dot f_i(\phi)
)
\right]
\right\} \ .
\end{equation}
The potential $U_\Omega$  is not positive definite, but is regular in the entire range $-\infty<x<\infty$.
Also, it vanishes at the BH event horizon  and at infinity.
It follows that eq. (\ref{stab6}) will have no bound
states if $U_\Omega$ is everywhere larger than the lower of its two asymptotic values, $i.e.$, if it is positive. 

For the case of the exponential, $\cosh$ and power-law coupling, the potential is, generically, everywhere positive for the vast majority of the  solutions analysed, which are therefore free of instabilities - see the related analysis in~\cite{Herdeiro:2018wub,Myung:2018jvi}. For the fractional coupling, on the other hand, there can be negative regions in the potential both for physical and exotic solutions. As an illustration,  in Fig.~\ref{F4b} the potential is plotted for a sequence of solutions.  One can see that the potential  is smaller than zero in a small $q$-region close to the RN limit  -- the RN BHs has the zero mode at $q=0.649$ ($\alpha=-10$). Then the potential becomes positive and remains so for arbitrary large $q$ along the remaining $\alpha$ branch. For the fractional coupling, on the other hand, there can be negative regions in the potential both for physical and exotic solutions.We emphasise that the existence of a negative potential region is a necessary, but not sufficient, condition for instability. It would be interesting to see if one can establish stability even in the presence of such negative regions, using, for instance the $S$-deformation method~\cite{Kimura:2017uor,Kimura:2018whv}.

		\begin{figure}[h]
	 \centering
	  \includegraphics[scale=0.4]{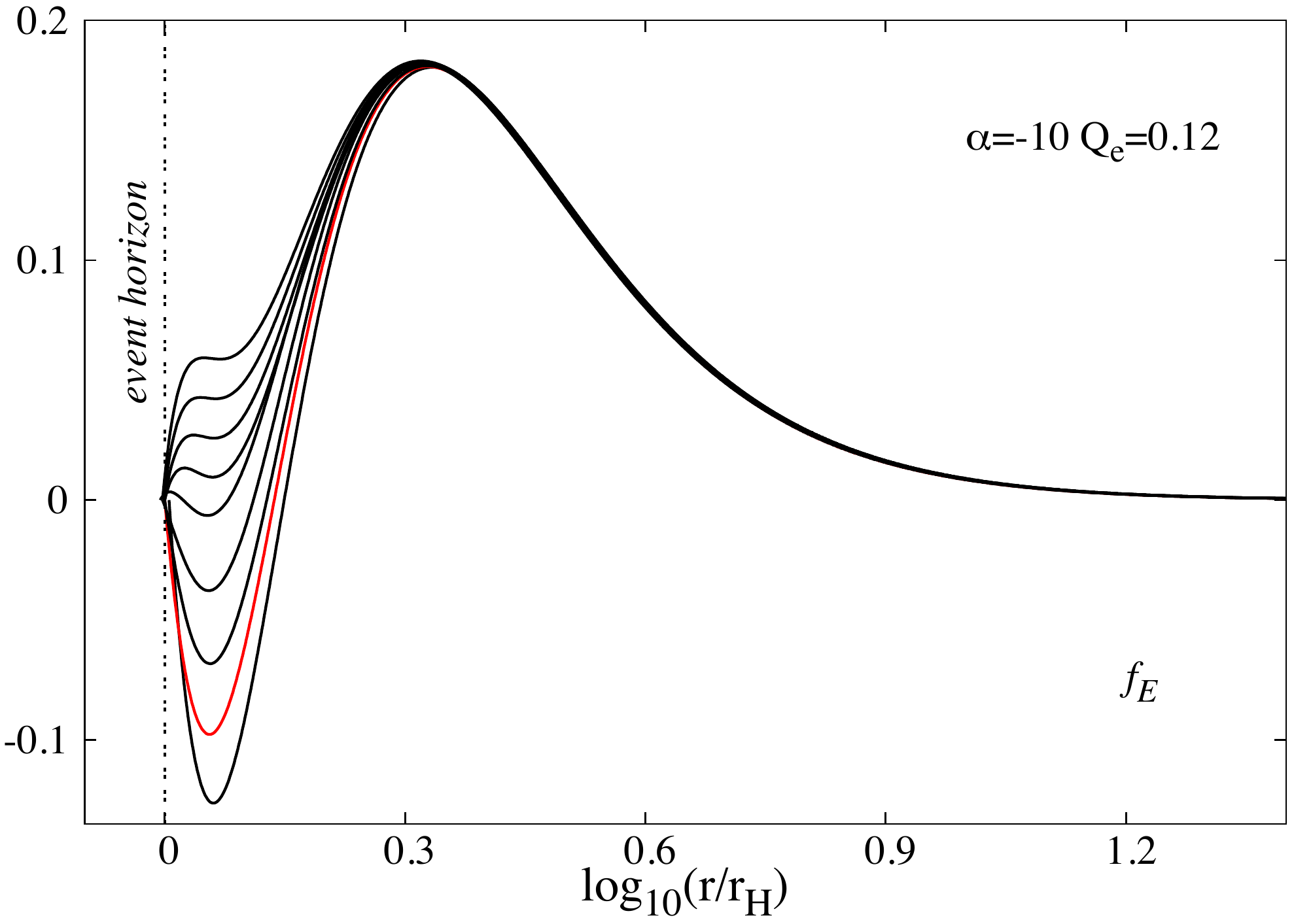}
	  \caption{Effective potential, $U_\Omega$, for a sequence of solution with the exponential coupling, $\alpha=-10$ and $Q_e=0.12$. The solutions have $r_H=0.32$ ($q=0.658$) -- lowest curve -- up to $r_H=0.308$ ($q=0.676$) -- top curve. The curve in red corresponds to the $f_E$ solution in Fig.~\ref{F1} (top left panel) with $r_H=0.318$ ($q=0.66$).}
	 \label{F4b}
	\end{figure}

%
%%%%%%%%%%%%%%%%%%%%%%%%%%%%%%%%%%%%%%%%%%%%%%%%%%%%%%%%%%%%%%%%%%%%%%%%%%%%%%%%%%%%%%
%%%%%%%%%%%%%%%%%%%%%%%%%%%%%%%%%%%%%%%%%%%%%%%%%%%%%%%%%%%%%%%%%%%%%%%%%%%%%%%%%%%%%%
\section{Entropic preference}\label{S4}
%%%%%%%%%%%%%%%%%%%%%%%%%%%%%%%%%%%%%%%%%%%%%%%%%%%%%%%%%%%%%%%%%%%%%%%%%%%%%%%%%%%%%%
%%%%%%%%%%%%%%%%%%%%%%%%%%%%%%%%%%%%%%%%%%%%%%%%%%%%%%%%%%%%%%%%%%%%%%%%%%%%%%%%%%%%%%
%

In the EMS scalar model, the Bekeinstein-Hawking BH entropy formula holds. Thus, the entropy analysis reduces to the analysis of the horizon area. It is convenient to use the already introduced reduced event horizon area $[a_H \equiv A_H/(16\pi M^2)]$. Then,  in the region where the RN BH and scalarised BHs co-exist -- the non-uniqueness region --,  for the same $q$ the scalarised solutions are always \textit{entropically preferred}. This is shown in Fig.~\ref{F4} for all four coupling $f_i (\phi ) $ functions studied herein.
One also observes that, for the same $q$, $a_H$ increases with the growth of $|\alpha |\ze $. 

	Such entropic considerations are not, however, sufficient to establish if the endpoint of the instability of the RN BH is the corresponding hairy BH with the same $q$. In~\cite{Herdeiro:2018wub}, fully non-linear dynamical evolutions were performed that established that for $f_E $, and sufficiently small $q$, this is indeed the case, which is consistent with the observation above that the scalarised solutions for the exponential (and also power-law and hyperbolic) coupling are, generically, stable against spherical perturbations. The endpoint of the instability, however, can only be established once fully non-linear numerical evolutions are studied. Such evolutions will be addressed in the next section. 
	
	An intriguing question, however, concerns the fractional coupling. Fixing the coupling, there are RN BHs that are unstable against scalar perturbations above the existence line in Fig.~\ref{F3} (right panel). However, no scalarised BHs exist for that value of $q$ (because it is above the critical set), in the physical region of the domain of existence with positive energy density. 
	The endpoint of the instability of such RN BHs is therefore an interesting question.

		\begin{figure}[h]
	 \centering
	  \includegraphics[scale=0.4]{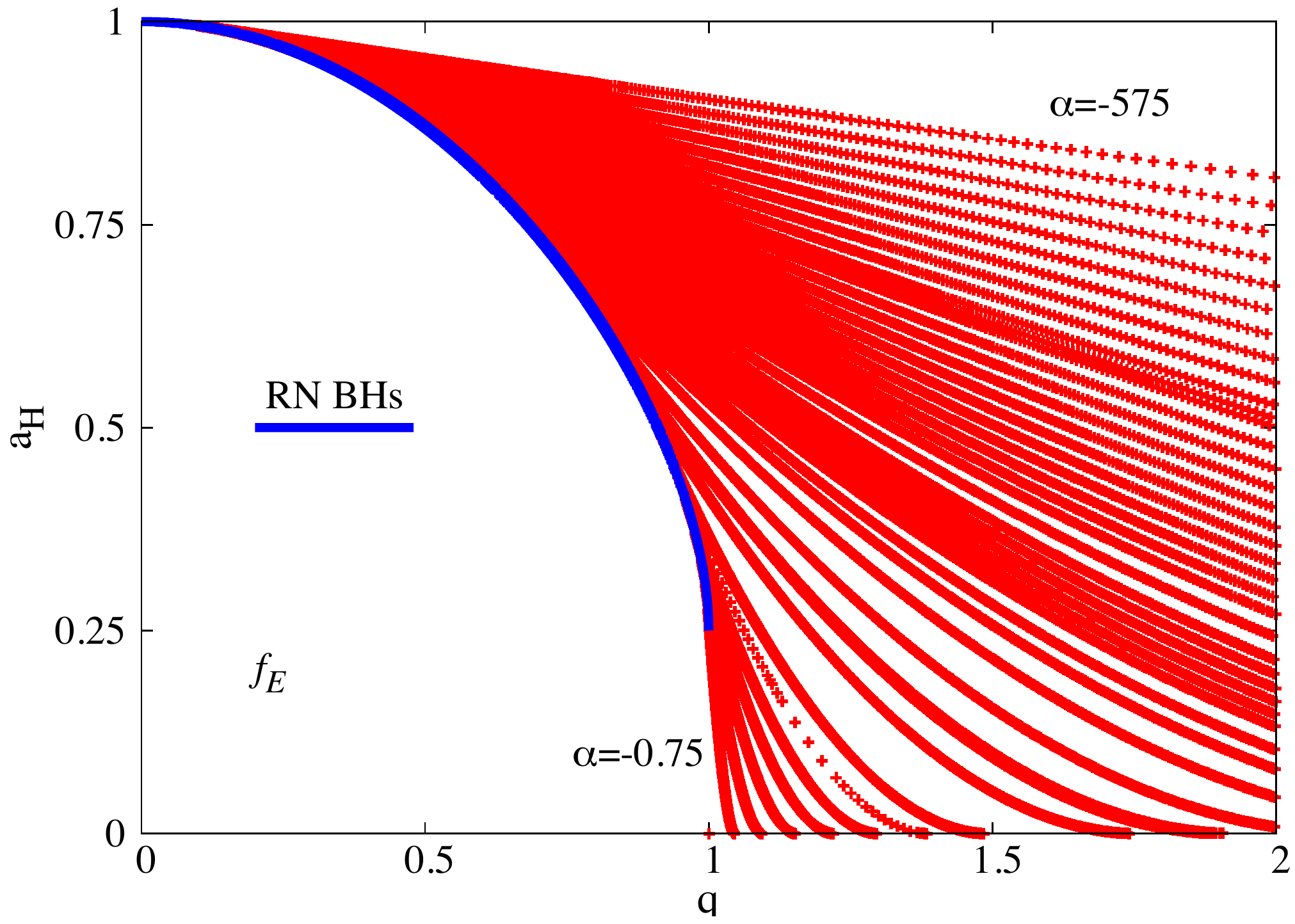}\hfill
	  \includegraphics[scale=0.4]{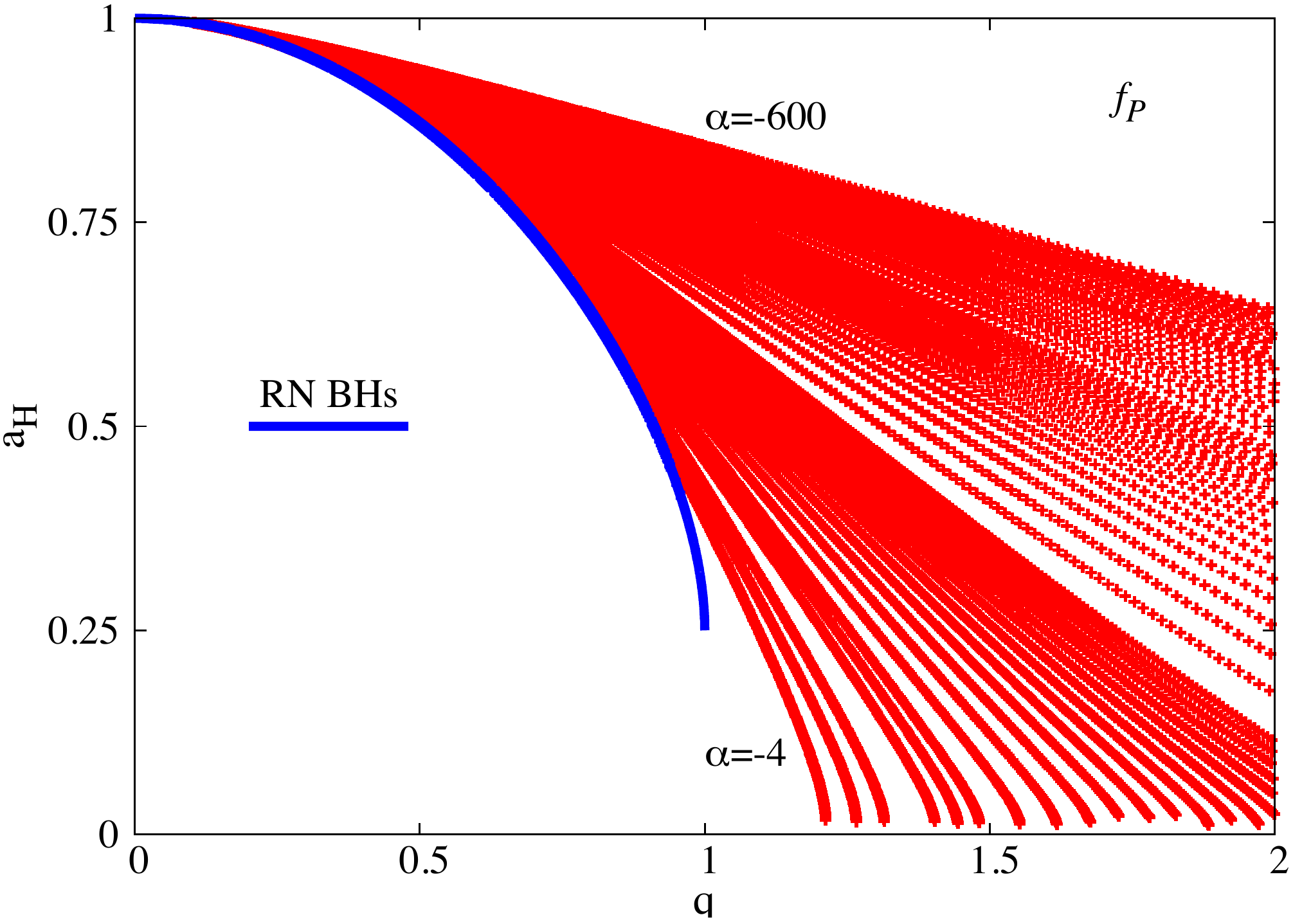}\\
	    \includegraphics[scale=0.4]{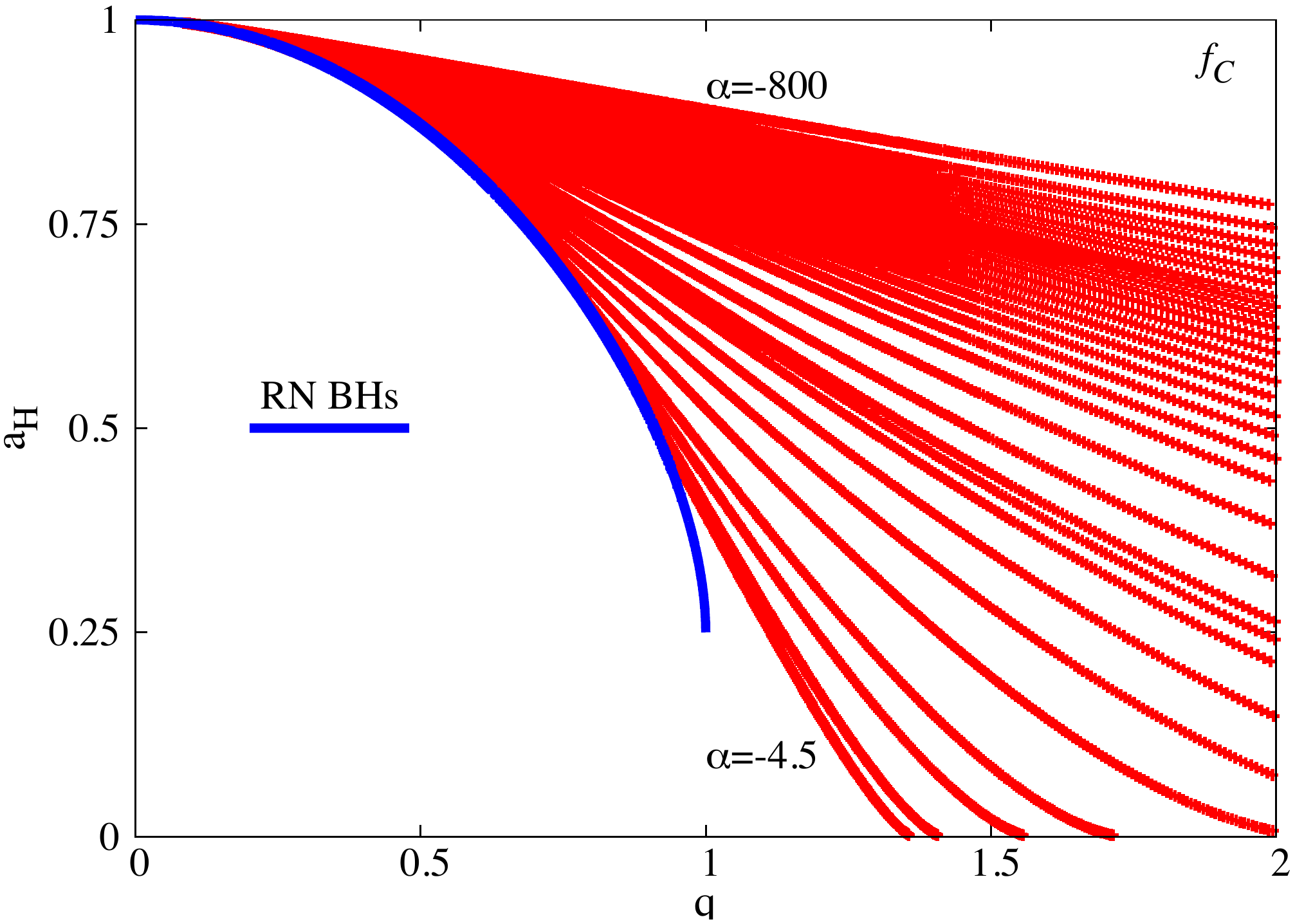}\hfill
	  \includegraphics[scale=0.4]{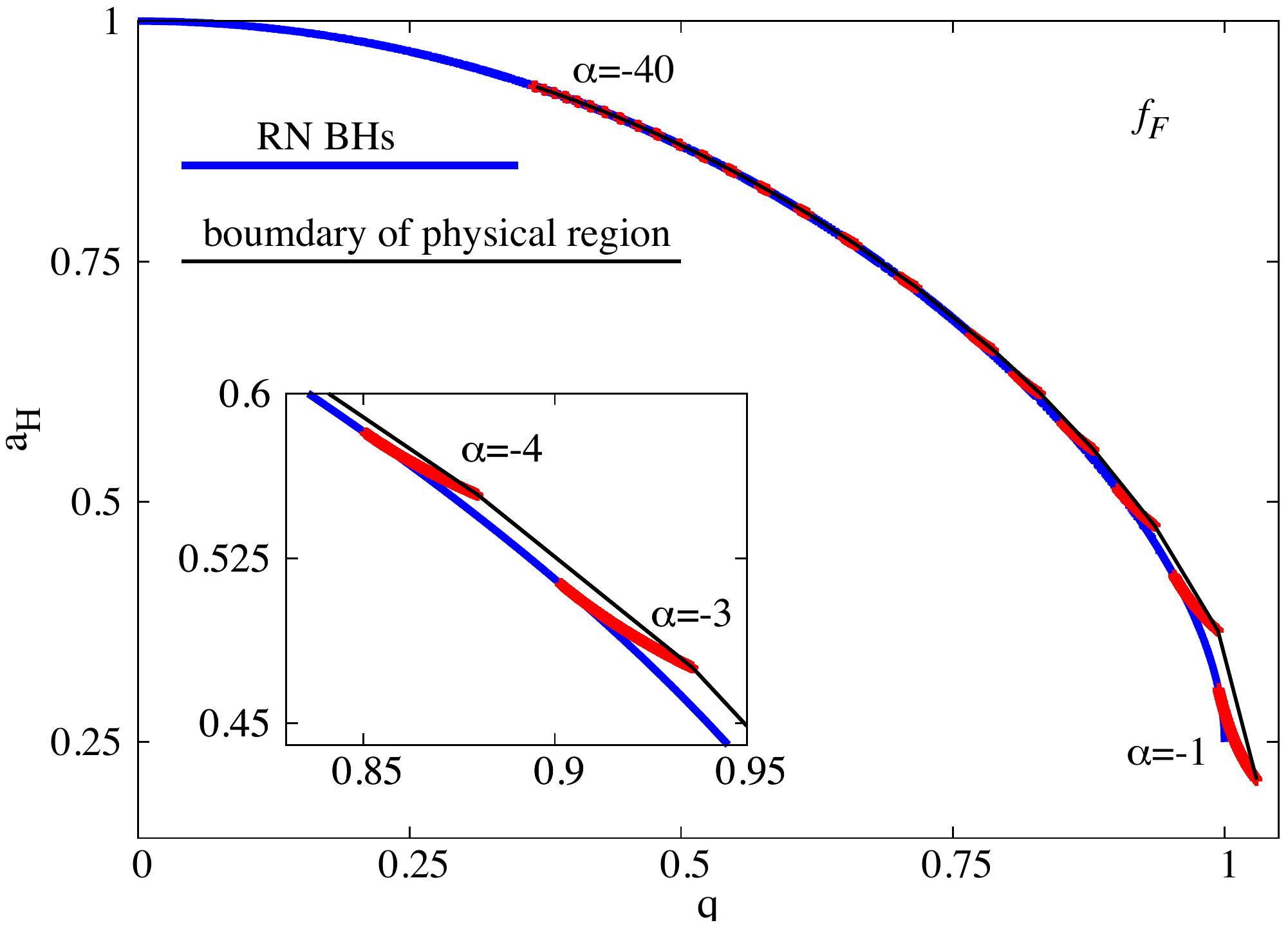}
	  \caption{Reduced area $a_H$ $vs.$ $q$ for: (top, left panel) $f_E\ze (\phi )\ze$; (top, right panel) $f_P \ze (\phi)\ze $; (bottom left panel) $f_C \ze (\phi) \ze$; (bottom right panel) $f_F \ze (\phi) \ze$. The blue lines are the sequence of non-scalarised RN BHs.
	 The red lines are sequences of (numerical data points representing) scalarised BHs for a given $\alpha$. Different sequences are presented, for a range of values of $\alpha\ze $. The solid black line shows the sequence of solutions along the boundary of the physical region for the $f_F\ze$ model.}
	 \label{F4}
	\end{figure}

    %%%%%%%%%%%%%%%%%%%%%%%%%%%%%%%%%%%%%%%%%%%%%%%%%%%%%%%%%%%%%%%%%%%%%%%%%%%%%%%%%%%%%%
%%%%%%%%%%%%%%%%%%%%%%%%%%%%%%%%%%%%%%%%%%%%%%%%%%%%%%%%%%%%%%%%%%%%%%%%%%%%%%%%%%%%%%
\section{Dynamical preference}\label{S5}
%%%%%%%%%%%%%%%%%%%%%%%%%%%%%%%%%%%%%%%%%%%%%%%%%%%%%%%%%%%%%%%%%%%%%%%%%%%%%%%%%%%%%%
%%%%%%%%%%%%%%%%%%%%%%%%%%%%%%%%%%%%%%%%%%%%%%%%%%%%%%%%%%%%%%%%%%%%%%%%%%%%%%%%%%%%%%
%
Following~\cite{Herdeiro:2018wub}, in which the numerical framework of~\cite{Sanchis-Gual:2015lje,Sanchis-Gual:2016tcm,Hirschmann:2017psw} was used, we have performed fully non-linear evolutions of unstable RN BHs in the EMS system under a small Gaussian scalar spherical perturbation, to assess the dynamical endpoint of the evolution. 

We have also considered evolutions with a non-spherical perturbation using the freely available \texttt{Einstein Toolkit}~\cite{toolkit2012open,Loffler:2011ay}. The scalar field initial data is 
\begin{equation}\label{initdata}
\phi(r,\theta) = A_{0}\,e^{-(r-r_0)^2/\lambda^2}\,Y^{0}_{\ell}(\theta)\ ,
\end{equation}
where $Y^{0}_{\ell}$ is the $\ell$-spherical harmonic with $m=0$ and $A_0,r_0$ two constants defining the amplitude and centre of the Gaussian radial profile of the scalar perturbation.

We implemented the Maxwell equations together with the evolution equations in~\cite{Hirschmann:2017psw} for a non-minimally coupled massless scalar field $\phi$ and an auxiliary variable, $\Pi\equiv -n^{\mu}\nabla_{\mu}\phi$, with $n^{\mu}$ the 4-velocity of the Eulerian observer. Two extra variables $\Psi_E$ and $\Phi_B$ are included to dynamically damp the constraints with two parameters, $\kappa_1$ and $\kappa_2$, which we take to be equal to 1. The set of evolution equations for an arbitrary coupling take the form
\begin{eqnarray}\label{evo_eq}
\bigl(\partial_t - {\cal L}_\beta \bigr) \phi & = & - \alpha_0 \Pi \ ,\\
\bigl(\partial_t - {\cal L}_\beta \bigr) \Pi & = & - D^a\bigl( \alpha_0 D_a \phi \bigr) + \alpha_0 K \Pi \nonumber\\
  & & \quad + \alpha_0 \, \frac{\dot{f}_{i}}{2} \bigl[ B_a B^a - E_a E^a \bigr] \  ,\\ 
\bigl( \partial_t - {\cal L}_\beta \bigr) E^a \! & = & \epsilon^{abc} D_b \bigl( \alpha_0 B_c \bigr) + \alpha \bigl[ K E^a - D^a \Psi_E \bigr] 
  \cr 
  & & \quad +  \alpha_0 \, \frac{\dot{f}_{i}}{f_{i}} \bigl[ \epsilon^{abc} D_b \phi B_c + \Pi E^a \bigr] \ , \\ 
\bigl( \partial_t - {\cal L}_\beta \bigr) \Psi_E & = & -\alpha_0 \bigl[ \frac{\dot{f}_{i}}{f_{i}} D_b \phi E^b - D_b E^b - \kappa_1 \Psi_E \bigr] \ , \\ 
\bigl( \partial_t - {\cal L}_\beta \bigr) B^a \! & = & \! - \epsilon^{abc} D_b \bigl( \alpha_0 E_c \bigr) + \alpha_0 \bigl[ K B^a + D^i \Phi_B \bigr] \  ,\\ 
\bigl( \partial_t - {\cal L}_\beta \bigr) \Phi_B & = & \alpha_0 \bigl[ D_b B^b - \kappa_2 \Phi_B \bigr] \ ,  
\end{eqnarray} 
where $\alpha_0$ is the lapse function, $\beta$ is the shift vector, $\gamma_{ij}$ are the 3-metric components, $D_a$ is the covariant derivative with respect to the 3-metric, and $E^{a}$ and $B^{a}$ are the electric and magnetic fields respectively. The matter source terms are given by

\begin{eqnarray} \label{sources}
\rho = n^{\alpha} n^\beta T_{\alpha\beta} &=& \frac{1}{8\pi}\bigl[D_a\phi D^a \phi + \Pi^2 + f_{i} \bigl( B_{a} B^{a} + E_a E^a \bigr) \bigl] \ , \\
j_a  = - n^\alpha \gamma_{a}^\beta T_{\alpha \beta} &=&   \frac{1}{4\pi}\bigl(- \Pi D_a\phi - f_{i} \epsilon_{abc} E^b B^c\bigl) \  ,\\  
S_{ab} = \gamma_{a}^\alpha \gamma_{b}^\beta T_{\alpha\beta} &= & \frac{1}{4\pi}\biggl[D_a \phi \, D_b \phi + f_{i} \, \bigl( B_a B_b  \, - E_a E_b  \bigr) \nonumber \\
  & & -   \frac{1}{2}\gamma_{ab} \, \Bigl[ D^c\phi \, D_c\phi - \Pi^2 + f_{i} \, \bigl( B_{c} B^{c} - E_c E^c \bigr) \Bigr]\biggl] \ .
\end{eqnarray} 

\begin{figure}[h]
	 \centering
	  \includegraphics[scale=0.45]{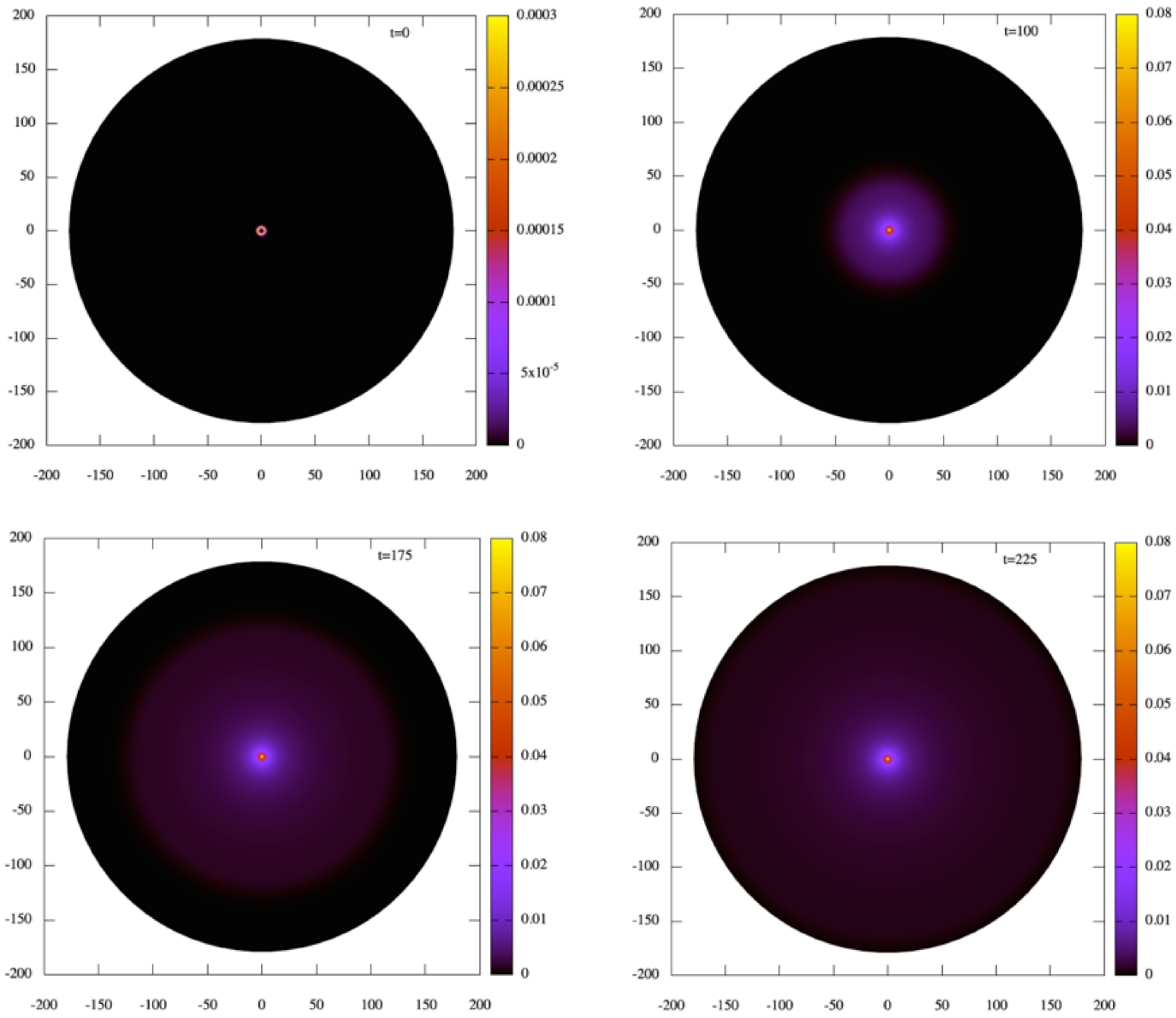}
	  \caption{Four snapshots of the time evolution of the scalar field around an unstable RN BH with $q=0.2$ in the EMS system, with the exponential coupling and  $\alpha=-400.979$.}
	 \label{F5}
	\end{figure}

To perform the evolutions we have used a numerical grid with 11 refinement levels with 
\[ \lbrace(192, 96, 48, 24, 12, 6, 3, 1.5, 0.75, 0.375, 0.1875) \ ,  \ \ (6.4, 3.2, 1.6, 0.8, 0.4, 0.2, 0.1, 0.05, 0.025, 0.0125, 0.00625)\rbrace \ , \]  where the first set of numbers indicates the spatial domain of each level and the second set indicates the resolution. Due to the geometry of the spherical harmonics, we consider equatorial-plane symmetry and reflection symmetry with respect to the $x$-$z$ plane for the $(\ell=2, m=0)$, but not for the $(\ell=1, m=0)$ mode, and reflection symmetry with respect to the positive values of $x$ and $y$ for both modes.

In~\cite{Herdeiro:2018wub} the dynamical formation of scalarised BHs with the exponential coupling was established. The evolution of the process can be observed in  Fig.~\ref{F5}, wherein four snapshots, at times $t=0,100,175,225$, are show for the exponential coupling, $q=0.2$ and $\alpha=-400.979$. The $\ell=0$ small Gaussian perturbation triggered the growth of a scalar cloud in the vicinity of the horizon that expands outwards and becomes a monotonically decreasing function of the radial coordinate. The energy transfer to the scalar field saturates by $t\sim 100$~\cite{Herdeiro:2018wub} and it reaches an equilibrium state, at least in the vicinity of the BH, around $t\sim 200$, albeit part of the more exterior scalar field distribution is still evolving outwards, settling down to the scalarised solution. The same qualitative pattern is observed for other couplings for which scalarisation occurs.

The endpoint of the evolution shown in Fig.~\ref{F5} is a scalarised BH with the same value of $q$. This was established by comparing the value of the scalar field on the horizon obtained in the numerical evolution with the one of the previously computed static scalarised solution with the same coupling and $q$. As explained above, fixing $\alpha$ the value of $\phi_0\equiv \phi(r_H)$ serves as a measure of $q$. In Fig.~\ref{F6} (left panel) this comparison is made for various values of $\alpha$, fixing $q=0.2$ of the initial RN BH, for both the exponential coupling (data already shown in~\cite{Herdeiro:2018wub}) and the power law coupling. The crosses are from the numerical evolutions and the solid line from the static solutions. The agreement is quite good. As discussed above, the power-law coupling produces a weaker scalarisation for the same coupling.

Fig.~\ref{F6} (right panel) performs a similar comparison, for the exponential coupling, but now exploring a larger range of values of $q$. Beyond $q\sim 0.4$, the agreement between the value of the scalar field on the horizon obtained from the evolutions and that obtained from the static solutions with the same $q$, ceases to hold. In other words, the endpoint of the evolution of a RN BH with a certain value of $q$ is not a scalarised BH with the same value of $q$. Rather, the former matches a scalarised BH with a lower value of $q$. This is interpreted as a non-conservative evolution which ejects a larger fraction of electric charge than energy when forming the scalarised BH. 

\begin{figure}[h]
	 \centering
	  \includegraphics[scale=0.3]{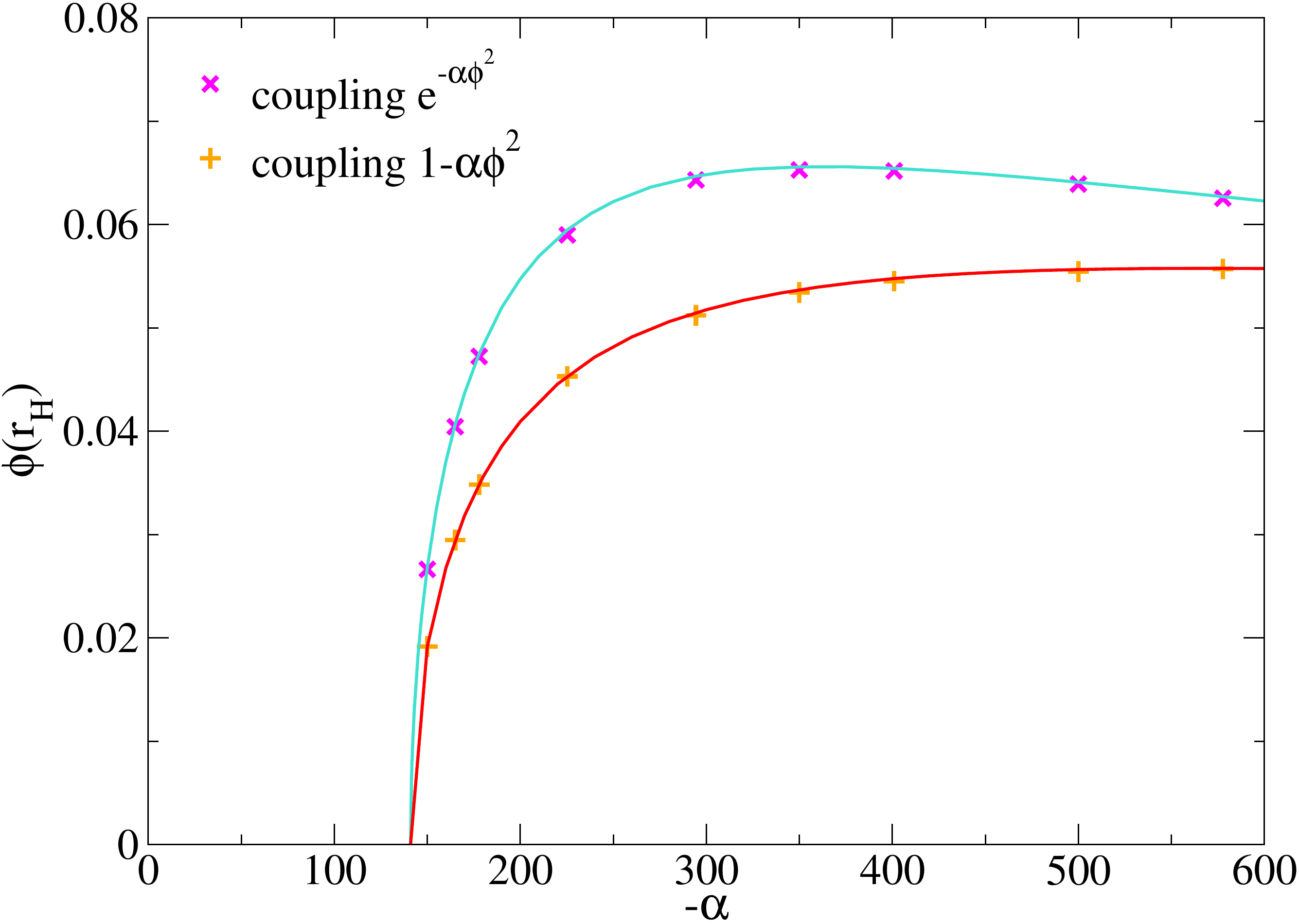}
	    \includegraphics[scale=0.3]{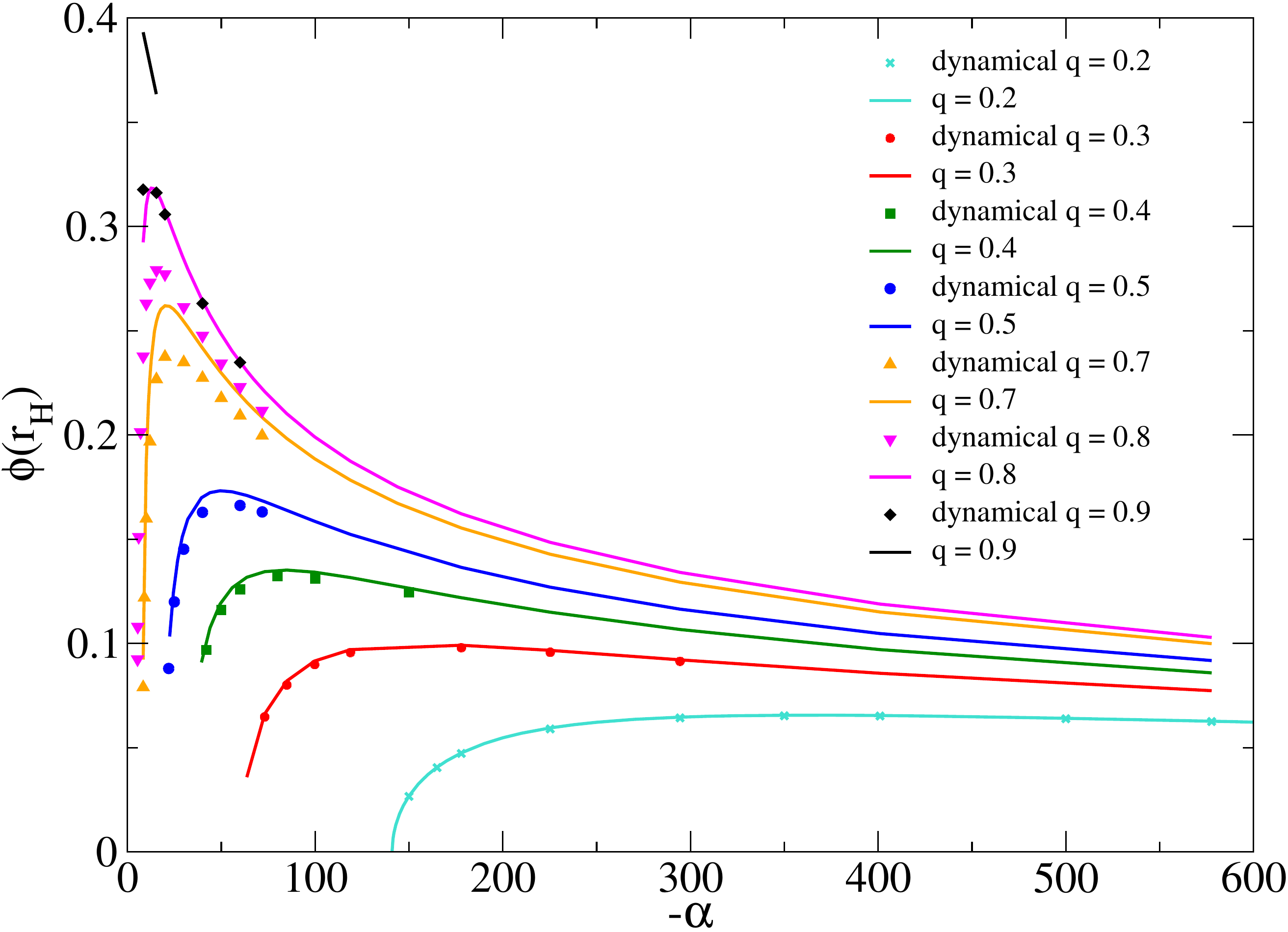}
	  \caption{(Left panel) Scalar field value at the horizon for $q=0.2$ and a range of couplings $\alpha$, for the exponential and power-law coupling. The solid line is obtained from the static solutions. The crosses are the dynamically obtained value from the numerical simuations after saturation and equilibrium has been reached. The agreement is notorious. (Right panel) A similar study, for the exponential coupling, but for various values of $q$. The agreement between the points and the lines with the same $q$ is restricted to $q\lesssim 0.4$. For larger $q$, the evolution points match static solution lines with a smaller $q$.}
	 \label{F6}
	\end{figure}

An intriguing possibility raised in~\cite{Herdeiro:2018wub} concerns the dynamical role of \textit{non-spherically symmetric} scalarised solutions. To address this issue we have performed the evolutions of an unstable RN BH under a non-spherical perturbations, using~\eqref{initdata} with $\ell=1,2$. In Fig.~\ref{F7} we show snapshots of such an evolution for the $\ell=2$ case. It can be observed that, initially, the non-spherical mode dissipates/is absorbed; then scalarisation proceeds much as in the case of a spherical perturbation. Similar results are obtained for the $\ell=1$ perturbation. Thus, scalarisation is robust, even without imposing spherical symmetry and, moreover, we see no evidence of the formation of the non-spherical scalarised solutions described in~\cite{Herdeiro:2018wub}. This suggests such solutions may be unstable.

    \begin{figure}[h]
	 \centering
	  \includegraphics[scale=0.45]{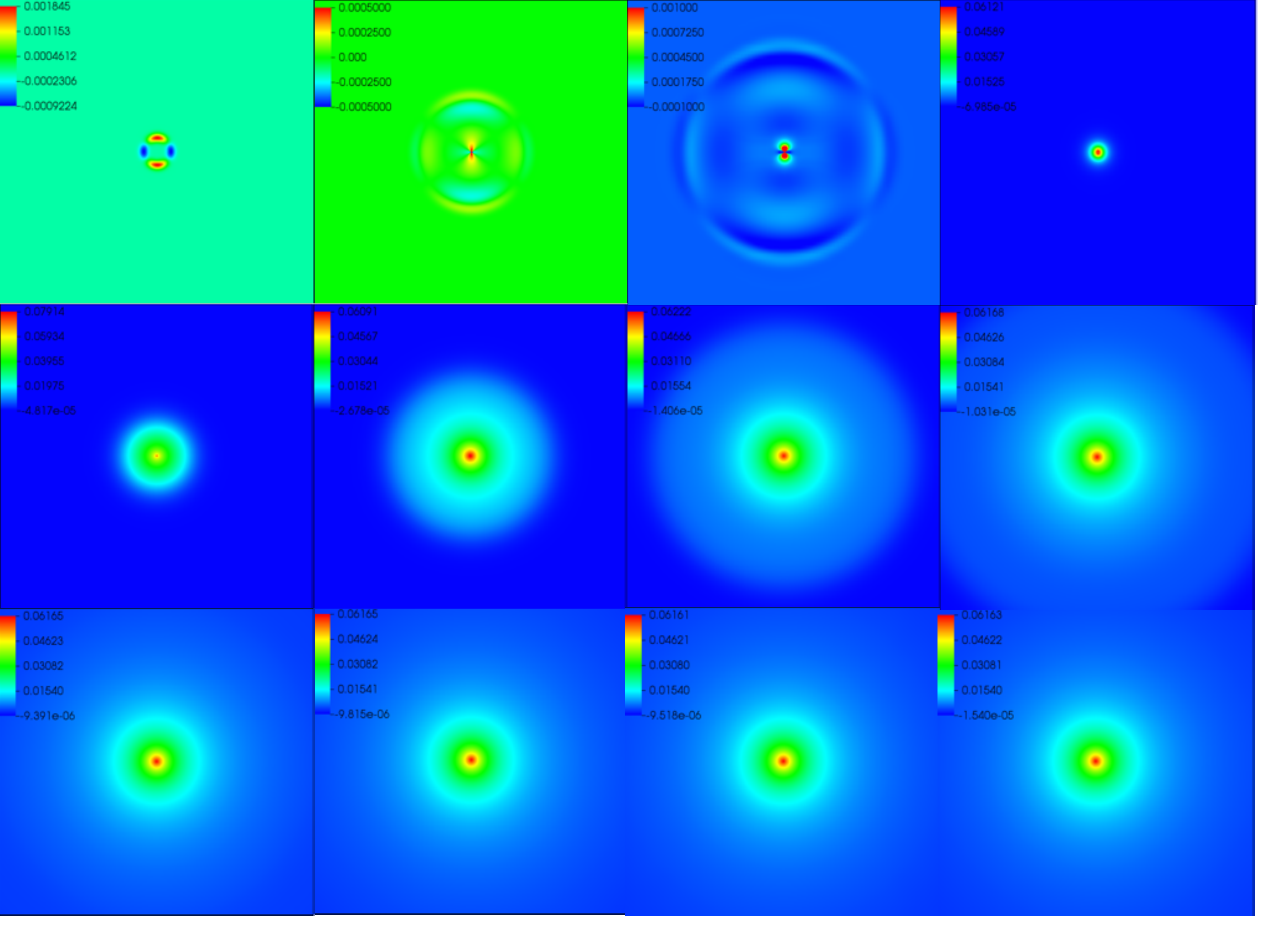}
	  \caption{Twelve snapshots in the $x-z$ ($y=0$) plane of the time evolution of an unstable RN BH with $q=0.2$ in the EMS system, with the exponential coupling and  $\alpha=-1200$ and an $\ell=2$, $m=0$ perturbation. The snapshots correspond to $t$  between $0$ and $140.8$. The data for negatives values of $x$ and $z$ are mirrored by the corresponding positive values, due to equatorial symmetry.}
	 \label{F7}
	\end{figure}

%
%%%%%%%%%%%%%%%%%%%%%%%%%%%%%%%%%%%%%%%%%%%%%%%%%%%%%%%%%%%%%%%%%%%%%%%%%%%%%%%%%%%%%%
%%%%%%%%%%%%%%%%%%%%%%%%%%%%%%%%%%%%%%%%%%%%%%%%%%%%%%%%%%%%%%%%%%%%%%%%%%%%%%%%%%%%%%
\section{Conclusions and remarks}\label{S6}
%%%%%%%%%%%%%%%%%%%%%%%%%%%%%%%%%%%%%%%%%%%%%%%%%%%%%%%%%%%%%%%%%%%%%%%%%%%%%%%%%%%%%%
%%%%%%%%%%%%%%%%%%%%%%%%%%%%%%%%%%%%%%%%%%%%%%%%%%%%%%%%%%%%%%%%%%%%%%%%%%%%%%%%%%%%%%
%

In this work we have studied BH scalarisation in the EMS model~\cite{Herdeiro:2018wub}, for four different choices of coupling function.  

Concerning the examination of the static solutions, two main conclusions can be extracted from our study. Firstly, for all cases studied, the scalarised solutions are entropically favoured over a comparable RN BH in the region where non-uniqueness holds. This creates a difference with the case of BH scalarisation in eSTGB model, where for the same power-law coupling we have considered here, the scalarised BHs are not entropically favoured and the scalarised spherically symmetric, fundamental BH solutions are not necessarily perturbative stable. Thus, BH scalarisation in the EMS and eSTGB models do not necessarily mimick one another, for all couplings. 
Secondly, the power-law, hyperbolic and exponential coupling are qualitatively very similar, albeit the exponential coupling maximises differences with respect to the RN case. The fractional coupling, on the other hand, yields qualitative differences with the existence of a different type of boundary in the domain of existence, bounding the region where physical solutions exist, abiding the weak energy condition. This boundary is associated to the divergent behaviour of the coupling for a certain value of the scalar field.

Concerning the dynamical evolutions, we have established that for small values of $q$ the evolutions of unstable RN BH lead to the formation of a scalarised BH with the same value of $q$, within numerical error. The evolution is essentially conservative. This was observed for the exponential and power-law coupling explicitly. Although we have not done evolutions with the hyperbolic coupling, it is very likely the same is observed. But for sufficiently high values of $q$ scalarisation decreases this value, thus establishing a non-conservative process is taking over, expelling from the BH a non-negligible fraction of charge and energy, with a dominance of the former. We have studied this in detail in the exponential coupling case, but expect the same result to be observed in the power-law and hyperbolic coupling. For the case of the fractional coupling, we have only performed evolutions at large $q$ and in the region where RN BHs overlap with (physical) scalarised BHs. Scalarisation was observed and a decrease in the value of $q$ occurred. Finally, we have analysed the evolution of unstable RN BHs under non-spherical perturbations and observed that a spherical scalarised BH emerges.

As an avenue of further research one may include a mass term for the scalar field. As in the case of other scalar-tensor theories this is expected to suppress the effects of scalarisation. We have done preliminary results of this model and observed that: 1) the existence line changes; 2) scalarisation requires a larger $|\alpha|$ as compared to the scalar-free case; and 3) the mass term quenches the dispersion of the scalar field, which becomes more concentrated in the neighbourhood of the horizon.  It would be interesting to analyse such inclusion of a mass term in greater detail.

\medskip
%
%%%%%%%%%%%%%%%%%%%%%%%%%%%  
\section*{Acknowledgements}
%%%%%%%%%%%%%%%%%%%%%%%%%%%
%
This work has been supported by Funda\c{c}\~ao para a Ci\^encia e a Tecnologia (FCT),
within project UID/MAT/04106/2019 (CIDMA), by CENTRA (FCT) strategic project UID/FIS/00099/2013, by national funds (OE), through FCT, I.P., in the scope of the framework contract foreseen in the numbers 4, 5 and 6
of the article 23, of the Decree-Law 57/2016, of August 29,
changed by Law 57/2017, of July 19. NSG  is supported by an FCT post-doctoral grant through the project PTDC/FIS-OUT/28407/2017 and A. Pombo is supported by the FCT grant PD/BD/142842/2018.   This work has further been supported by  the  European  Union's  Horizon  2020  research  and  innovation  (RISE) programmes H2020-MSCA-RISE-2015
Grant No.~StronGrHEP-690904 and H2020-MSCA-RISE-2017 Grant No.~FunFiCO-777740. The authors would like to acknowledge networking support by the
COST Action CA16104.

%%%%%%%%%%%%%%%%%%%%%%
%%%   REFERENCES   %%%
%%%%%%%%%%%%%%%%%%%%%%

  \bibliographystyle{ieeetr}
  \bibliography{biblio}

%%%%%%%%%%%%%%%
%%%   END   %%%
%%%%%%%%%%%%%%%

\end{document}